\documentclass[twocolumn]{aastex63}
\usepackage[]{units}
\usepackage[update,prepend]{epstopdf}
\usepackage{graphicx}
\usepackage{amssymb}
\usepackage{amsmath}
\usepackage{threeparttable}
\usepackage{hyperref}
\usepackage{color}

\usepackage{mathtools}

\usepackage{makecell}

\newcommand{\kms} {\,km\,s$^{-1}$}

\newcommand{\ang}{\,\AA}

\mathchardef\mhyphen="2D
\shorttitle{Metallicity and age dependence of wide binaries}
\shortauthors{Hwang et al.}

\graphicspath {{figs/} }

\begin{document}

\title{The non-monotonic, strong metallicity dependence of the wide-binary fraction}

\author[0000-0003-4250-4437]{Hsiang-Chih Hwang}
\affiliation{Department of Physics \& Astronomy, Johns Hopkins University, Baltimore, MD 21218, USA}
\author[0000-0001-5082-9536]{Yuan-Sen Ting}
\thanks{Hubble Fellow}
\affiliation{Institute for Advanced Study, Princeton, NJ 08540, USA}
\affiliation{Department of Astrophysical Sciences, Princeton University, Princeton, NJ 08544, USA}
\affiliation{Observatories of the Carnegie Institution of Washington, 813 Santa Barbara Street, Pasadena, CA 91101, USA}
\affiliation{Research School of Astronomy \& Astrophysics, Australian National University, Cotter Rd., Weston, ACT 2611, Australia}

\author[0000-0001-5761-6779]{Kevin C. Schlaufman}
\affiliation{Department of Physics \& Astronomy, Johns Hopkins University, Baltimore, MD 21218, USA}
\author[0000-0001-6100-6869]{Nadia L. Zakamska}
\affiliation{Department of Physics \& Astronomy, Johns Hopkins University, Baltimore, MD 21218, USA}

\author[0000-0002-4013-1799]{Rosemary F.G. Wyse}
\affiliation{Department of Physics \& Astronomy, Johns Hopkins University, Baltimore, MD 21218, USA}
\begin{abstract}

The metallicity dependence of the wide-binary fraction in stellar populations plays a critical role in resolving the open question of wide binary formation. In this paper, we investigate the metallicity ([Fe/H]) and age dependence of the wide-binary fraction (binary separations between $10^3$ and $10^4$\,AU) for field F and G dwarfs within 500\,pc by combining their metallicity and radial velocity measurements from LAMOST DR5 with the astrometric information from Gaia DR2. We show that the wide-binary fraction strongly depends on the metallicity: as metallicity increases, the wide-binary fraction first increases, peaks at [Fe/H]$\simeq 0$, and then decreases at the high metallicity end. The wide-binary fraction at [Fe/H]$=0$ is about two times larger than that at [Fe/H]$=-1$ and [Fe/H]$=+0.5$. This metallicity dependence is dominated by the thin-disk stars. Using stellar kinematics as a proxy of stellar age, we show that younger stars have a higher wide-binary fraction at fixed metallicity close to solar. We propose that multiple formation channels are responsible for the metallicity and age dependence. In particular, the positive metallicity correlation at [Fe/H]$<0$ and the age dependence may be due to the denser formation environments and higher-mass clusters at earlier times. The negative metallicity correlation at [Fe/H]$>0$ can be inherited from the similar metallicity dependence of close binaries, and radial migration may play a role in enhancing the wide-binary fraction around the solar metallicity.

\end{abstract}

\section{Introduction}

Wide binaries are weakly bound, as such they are sensitive to the gravitational perturbations in the Milky Way and have been used to investigate the visible and invisible Galactic structures \citep{Heggie1975,Bahcall1981,Bahcall1985,Weinberg1987,Chaname2004, Yoo2004,Quinn2009a,Jiang2010a}. Wide binaries may also be able to probe the dark matter substructure in dwarf galaxies \citep{Penarrubia2016}. Furthermore, a significant fraction of stars are in binaries and multiple systems \citep{Abt1976,Duquennoy1991,Fischer1992,Duchene2013}, and about half of wide binaries (separations $a>1000$\,AU) are the outer binaries of high-order hierarchical systems \citep{Raghavan2010,Tokovinin2014a, Tokovinin2014b,Moe2017}, so understanding the formation of wide binaries is crucial for the formation of hierarchical systems and the implications for both large-scale and small-scale Galactic structures.

The formation of wide binaries is still not well understood. They are unlikely to form by capture of random field stars, due to the low stellar density in the field (e.g. \citealt{Goodman1993}). This is further supported by the similarity of the chemical compositions of the wide binary components with separations $\lesssim1$ pc$\sim2\times10^{5}$\,AU \citep{Andrews2018, Andrews2019, Hawkins2020}, indicating that the components of wide binaries are born together, and several mechanisms have been proposed for their formation. For example, turbulent core fragmentation can form binaries with separations from $\sim100$\,AU to $\sim1000$\,AU \citep{Padoan2002,Fisher2004,Offner2010}. Binaries with separations of $10^3-10^5$\,AU can be formed through the dynamical unfolding of compact triples \citep{Reipurth2012}, the dissolution of star clusters \citep{Kouwenhoven2010, Moeckel2011}, or by the random pairing of adjacent pre-stellar cores \citep{Tokovinin2017}. 

Many observational efforts have been directed at constraining the formation of wide binaries. Several young ($<$ a few Myr) wide binaries have been found (e.g. \citealt{Kraus2011, Pineda2015, Tobin2016a,Lee2017}), supporting the proposal that wide binaries can be formed during the pre-main sequence phase, through turbulent core fragmentation and/or the pairing of pre-stellar cores. However, it is known that the wide-binary fraction is higher in pre-main sequence stars compared to that of field stars \citep{Ghez1993,Kohler2000}. Furthermore, the separation distribution of binaries in low-density star-forming regions \citep{Simon1997,Kraus2009,Tobin2016, Joncour2017} is found to be flatter than that of main-sequence field binaries \citep{Tokovinin2012,El-Badry2018b}. It has been argued that these differences may arise from the different formation environments and/or ages in the current young star-forming regions compared to those of the field stars \citep{Kroupa1995, Kraus2009}. Therefore, it remains challenging to directly infer the formation of field wide binaries from the multiplicity in young star-forming regions alone.

Theory has suggested that wide binaries can form from the dissolution of clusters \citep{Kouwenhoven2010, Moeckel2011}, which similar to the turbulent core fragmentation and the random pairing of pre-stellar cores, is also environment-dependent. In this scenario, wide binaries are formed by the pairing of initially unbound stars when the cluster rapidly expands after gas expulsion. The formation timescale of wide binaries in this case correlates with how fast the gas is dispersed, which is of the order of $\sim$10 Myr \citep{Lada2003,Bastian2005,Fall2005,Mengel2005}. \cite{Kouwenhoven2010} show that the dissolution of lower-mass clusters results in a higher wide-binary fraction because the lower velocity dispersion increases the pairing probability in the phase space. Observational studies find a lower wide-binary fraction in open clusters than that of the low-density star-forming regions and the field \citep{Bouvier1997,Deacon2020}. Since the surviving open clusters are usually at the massive end of the cluster mass function, these results are most likely due to that the high-density environments reduce the wide binary formation within the clusters. Therefore, the wide-binary fraction resulting directly from the cluster dissolution remains not well constrained.

Besides forming from the dissolution of clusters, wide binaries can also form through the dynamical unfolding of compact triples. Three stars are initially formed in a compact, unstable configuration, and the subsequent dynamical evolution can bring one component closer and push the other component further away, and if it is not ejected entirely, the object appears as a wide binary \citep{Reipurth2012}. This scenario is supported by that these wide systems are frequently members of higher-order multiples \citep{Law2010,Allen2012,Elliott2016}, and that main-sequence contact binaries have a factor of $3$ higher wide companion fraction at separations $>1000$\,AU than that of the field stars \citep{Hwang2020c}. However, the importance of the dynamical unfolding to form wide binaries is still debated. In particular, this scenario should lead to outer companions with eccentric orbits, but \cite{Tokovinin2020a} shows that the eccentricity distribution of wide binaries is close to thermal, inconsistent with the dynamical unfolding explanation.

With the multiple interconnected formation channels, the exact explanation remains elusive despite decades of research. This situation is drastically changing with the advent of recent large spectroscopic surveys. In particular, metallicity dependence of binary fraction measured from these surveys is proving useful in disentangling binary formation. In terms of close binaries, recent studies have shown that the close-binary fraction is anti-correlated with metallicity \citep{Grether2007, Raghavan2010,Yuan2015,Badenes2018,Moe2019,El-Badry2019a,Mazzola2020}. This finding supports the scenario that close binaries are formed via disk fragmentation and the lower-metallicity disks are more prone to fragmentation \citep{Kratter2006, Tanaka2014, Moe2019, Tokovinin2020}. Alternatively, radiation hydrodynamical simulations from \cite{Bate2019} suggest that the anti-correlation between close-binary fraction and metallicity could also be explained by the fact that lower metallicities facilitate all kinds of small-scale fragmentation (disk, filament, and core fragmentation), not just disk fragmentation. Regardless of their exact physical explanations, it has been proposed that such metallicity dependence of the close-binary fraction may ultimately be passed on to their final products -- blue stragglers \citep{Wyse2020}.

While the studies of close binaries have reached more consensus, the metallicity dependence of the wide-binary fraction is less conclusive, with competing conclusions from various studies \citep{ZapateroOsorio2004,Zinnecker2004, Rastegaev2008,Jao2009, Lodieu2009, Zhang2013, Ziegler2015}. In a recent study, \cite{El-Badry2019a} investigate the binary fraction as a function of separation and metallicity. They use Gaia DR2 to establish the comoving pair sample within 200\,pc \citep{El-Badry2018b}, and combine it with wide-field spectroscopic surveys, including LAMOST, RAVE, APOGEE, GALAH, and Hypatia. They find an emergence of an anti-correlation between the binary fraction and metallicity at separations $a<200$\,AU, while the binary fraction at $a\gtrsim200$\,AU remains relatively constant with respect to metallicity. The authors conclude that a significant fraction of binaries with $a<200$\,AU are formed by disk fragmentation while binaries with $a\gtrsim200$\,AU may be formed from turbulent core fragmentation which has a weaker dependence on metallicity. 

In this paper, we revisit the metallicity dependence of field wide binaries ($a=10^3-10^4$\,AU) using the LAMOST and Gaia DR2 datasets. Our sample consists of stars out to 500\,pc, important for inclusion of sufficient numbers of the thick-disk and halo stars. By doubling the sample size compared to \cite{El-Badry2019a} and, more importantly, dissecting the kinematics of these stars as age proxy which is left unexamined in \cite{El-Badry2019a}, we are able to investigate metallicity and age effects and better constrain both the formation and evolution scenarios for wide binaries.

Through out the paper, we refer to wide binaries as those with separations between $10^3$ to $10^4$\,AU. While we adopt the notation `binary' for our multiple systems, we caution that some of them might be actually unresolved higher-order multiples. The paper is organized as follows. Section~\ref{sec:sample} describes the LAMOST and Gaia datasets and the method of searching for comoving companions. Section~\ref{sec:result} presents our main result that the wide-binary fraction is strongly dependent on the metallicity. We discuss the implications for the wide binary formation in Section~\ref{sec:discussion} and conclude in Section~\ref{sec:conclusion}.

\section{Sample selection and method}
\label{sec:sample}
\subsection{LAMOST and metallicity measurements}
\label{sec:LAMOST}
Our sample is selected from The Large Sky Area Multi-Object Fiber Spectroscopic Telescope (LAMOST; \citealt{Deng2012,Zhao2012}). In its final data release of the LAMOST Phase I (2011–2017) survey, LAMOST DR5 has released optical ($3700-9000$\ang) low-resolution spectra ($R\sim1800$) for about 10 million stars, providing a rich dataset for Galactic archaeology.

We use the A/F/G/K stars catalog from LAMOST DR5 (v3). Its metallicity ([Fe/H]) is derived from the LAMOST Stellar Parameter pipeline (LASP; \citealt{Wu2011, Wu2011a}) and the data-driven Payne pipeline (DD-Payne; \citealt{Xiang2019}). LASP fits the observed spectrum using a full spectrum fitting package ULySS (Universit\'e de Lyon Spectroscopic analysis Software; \citealt{Koleva2009}). Specifically, each observed spectrum is fit to a grid of model spectra based on the ELODIE library \citep{Prugniel2001,Prugniel2007} to derive $T_{\rm eff}$, $\log g$, and [Fe/H].

The Payne is designed to measure multiple elemental abundances where the model spectra are emulated with neutral networks \citep{Ting2019}. When combined with data-driven models with theoretical prior, the Payne can be applied to low-resolution spectra ($R\sim1000$) to derive reliable elemental abundances, a method that has been dubbed the name Data-Driven Payne, or DD-Payne (\citealt{Ting2017, Ting2017a}, see also \citealt{Ness2015,OBriain2020}). Based on this, \cite{Xiang2019} train the spectral model using the LAMOST stars where the stellar labels have been derived from other high-resolution surveys like GALAH \citep{DeSilva2015} and APOGEE \citep{Majewski2017}. DD-Payne provides a final product with stellar parameters ($T_{\rm eff}$, $\log g$, $V_{mic}$) and 16 elemental abundances.

By comparing the metallicity measured by LASP and DD-Payne, we find that LASP metallicities are systematically higher than DD-Payne metallicities by 0.07 dex, with a standard deviation of 0.07 dex. The 0.07\,dex offset between LASP and DD-Payne does not strongly correlate with metallicity. Since LASP metallicities are calibrated to the ELODIE spectral library and DD-Payne ties the metallicity to APOGEE, this systematic offset might be due to the different abundance scale used by ELODIE and APOGEE (M. Xiang, private communication). Otherwise, the small scatter of 0.07\,dex shows that the metallicities of  LASP and DD-Payne are in good agreement. Since we focus on the relative trend of the wide-binary fraction with respect to metallicity, the accuracy of the absolute values of metallicities is not the main concern.

\subsection{Gaia and the comoving companion search}

We use Gaia Data Release 2 (DR2) to search for the comoving companions around the LAMOST stars. Gaia DR2 provides broad-filter G-band magnitudes, blue-band BP magnitudes, red-band RP magnitudes, sky positions, parallaxes, and proper motions for 1.33 billion objects and radial velocities for 7 million stars \citep{Gaia2016,Gaia2018Brown}, resulting in an unprecedented dataset for the phase-space information of the Milky Way.

We cross-match the LAMOST catalog with Gaia DR2 using a matching radius of 2\,arcsec. When an object has multiple visits by LAMOST, we only keep the one with the highest signal-to-noise ratio (SNR) in SDSS $g$-band. For LAMOST stars where multiple Gaia sources are matched, we keep the one with the smallest separation. If one Gaia \texttt{source\_id} is matched to multiple LAMOST designations, which most of the time is due to the repeated LAMOST designation naming, we keep the one with the highest SNR in SDSS $g$-band.

The comoving companion searching method is detailed in \cite{Hwang2020c}. Briefly, for each target star, we select a nearby star sample where the stars have similar parallaxes (either parallax difference $< 0.2$\,mas, Gaia DR2's typical parallax errors, or the line-of-sight distance difference $<20$\,pc) as the target star. Then between the target star and each nearby star, we compute the two-dimensional relative velocity (proper motion difference divided by the mean parallax of the two stars) on the sky and the projected separation. We do not consider the component along the line of sight because that involves the parallax difference of two stars, which is dominated by the parallax measurement errors. The physical projected distance remains accurate because it does not involve the parallax difference of the two stars. In the remaining paper, the relative velocity and separation refer to the projected quantities (except for the total velocity $v_{tot}$ explained in Sec.~\ref{sec:distinguish-pop} that uses the radial velocity component). The comoving companions are well separated from the chance projection stars in the relative velocity-separation space, and we use an empirical demarcation line introduced in \cite{Hwang2020c} to select comoving companions. 

To exclude contamination from open clusters or comoving groups, we exclude stars that belong to aggregates with more than 50 stars within separation $10^5-10^6$\,AU and relative velocity $<10$\kms. This only excludes 0.4\% of the targets and does not have a strong impact on our result. For the ease of interpretation and counting, we further exclude targets that have more than one comoving companion, which affects only 0.1\% of the sample.

\subsection{Selection criteria for the main sample}
\label{sec:main-selection}

To ensure that the LAMOST pipeline metallicity (LASP) are reliable for our targets, we require that their spectral SNR per pixel be $>$50 in the SDSS $g$-band. For DD-Payne metallicity, we require that their spectral SNR per pixel $>50$ in the SDSS $g$-band and the fitting quality flag \texttt{QFLAG\_CHI2==good}. We limit our sample to the effective temperature between 5000 and 7000\,K and surface gravity $\log g>3.7$. For DD-Payne, we require that  \texttt{TEFF\_FLAG==good} and \texttt{LOGG\_FLAG==good}. Based on the spectral type classified by LASP, $94\%$ of the selected sample are F and G dwarfs.

After cross-matching with Gaia DR2, we limit our sample to parallaxes $>2$\,mas (distances within 500\,pc). We further exclude unreliable photometric and astrometric measurements following the criteria in the Appendix B in \cite{Gaia2018Babusiaux}, except that we relax the criteria on BP and RP fluxes. These criteria require that the S/N of Gaia G band larger than 50, the parallax over error $>10$, the visibility periods $>8$, and an astrometric quality criteria introduced in the Appendix C of \cite{Lindegren2018}. We do not apply any criteria for BP and RP fluxes because BP and RP photometry has a worse spatial resolution than G band \citep{Evans2018}, and also BP and RP have inferior sensitivity than G-band, which may affect the detection of faint companions. We use the same criteria for the nearby star sample where the comoving companion is searched.

With these selections and the removal of sources in comoving groups, we end up with 257,560 stars with LASP metallicity, and 247,669 with DD-Payne metallicity. They share 243,823 stars in common.  LASP and DD-Payne are essentially the same sample of stars with two alternative [Fe/H] determinations. Tables of these wide binaries are available electronically, and their information is detailed in Appendix~\ref{sec:appendix-catalog}.

\subsection{Computing the wide-binary fraction}

In this study, we adopt a conservative angular resolution of 2\,arcsec, corresponding to $1000$\,AU at 500\,pc, i.e., we consider only wide binaries with separations of two stars to be at least 1000\,AU. This choice is motivated by several factors. Gaia G-band uses PSF-fitting photometry, and its spatial resolution is $\sim0.5$\,arcsec in DR2 \citep{Arenou2018}. Furthermore, we find that the number of comoving pairs decreases at angular separations $\lesssim1.5$\,arcsec, which may be due to the worse quality of astrometric measurements in the presence of a nearby source. Also, the diameter of LAMOST fibers is 3.3\,arcsec \citep{Zhao2012}. Therefore, two stars with an angular separations $\lesssim3.3/2$\,arcsec would be located in a single fiber, which may affect metallicity measurements.

We define the wide-binary fraction (WBF) to be
\begin{equation}
WBF = N_{\rm companion} / N_{\rm LAMOST},
\end{equation}
where $N_{\rm LAMOST}$ is the number of LAMOST stars, and $N_{\rm companion}$ is the number of LAMOST stars that have one comoving companion in Gaia at $10^3$ to $10^4$\,AU. 

In most cases, LAMOST does not observe both stars due to its random subsampling. The random subsampling does not affect our definition of wide binaries since we only require one of two stars to have LAMOST observations (and both stars have Gaia phase space information). Nonetheless, the metallicity of the two stars, individually, might not be accessible. Here we assume that both stars have the same metallicity and adopt the metallicity from the stellar component with LAMOST measurements. We argue that this assumption is justified because previous studies have shown that wide binaries with separations $\lesssim10^4$\,AU have nearly identical elemental abundances \citep{Andrews2018,Andrews2019, Hawkins2020}, and simulations have suggested that most pairs with separations $\lesssim 10^6$\,AU and small relative velocities ($\lesssim 2\,$km/s) are conatal \citep{Kamdar2020}. Due to turbulent mixing, conatal stars from the same gas cloud are expected to be homogeneous in metallicity \citep{Feng2014}.

One possible bias is the higher detection rate of fainter companions for less distant targets. Therefore, when computing $N_{\rm companion}$, we only consider companions that have absolute G-band magnitudes $<10$, where our companion search is complete across the entire distance range of the sample. This criterion removes most of white dwarf companions, which may induce an age dependence of $N_{\rm companion}$ if young, bright white dwarfs are detected but old, faint white dwarfs are not. This is not a strong effect because white dwarf-main sequence pairs are more than ten times less frequent than main sequence-main sequence pairs \citep{El-Badry2018b}. The absolute magnitude criterion also excludes faint M dwarf companions, but because the lifetime of M dwarfs is longer than the age of Universe, this does not induce age dependence. 

We select a sub-sample within 100\,pc to test the completeness. In this sub-sample, without the absolute magnitude constraint for the companions, the wide-binary fraction is $7.12\pm0.70$\%. This is consistent with \cite{Raghavan2010}  where they measure that $7\pm1$\% of solar-like stars within 25\,pc have companions at separations between 10$^3$ to 10$^4$\,AU. With an additional cut on absolute G-band magnitudes $<10$\,mag, the wide-binary fraction of our 100-pc sample is reduced to $3.73\pm0.51$\%, where 84\% of the excluded companions are faint M dwarfs and $16$\% are white dwarfs. The wide-binary fraction (with the absolute magnitude cut on the companions) of our full 500-pc sample is $2.98\pm0.03$, in good agreement with the 100-pc sample (1.5\,$\sigma$). This illustrates that the companions with absolute G-band magnitudes $<10$\,mag are well detected within 500\,pc. The 1.5\,$\sigma$ difference may arise from the different metallicity regime probed at larger distances and the slightly reduced completeness of Gaia sources at angular separations close to 2\,arcsec \citep{Arenou2018}. In the Appendix, we test with larger binary separations and show that our results are robust against the possible incompleteness at small angular separations.

Values of $N_{\rm LAMOST}$ and $N_{\rm companion}$ may weakly depend on the distance because of the spatial resolution. For example, in the case of triple stars, the counting of $N_{\rm LAMOST}$ and  $N_{\rm companion}$ is different depending on whether the inner binary of a hierarchical triple is resolved or not. Specifically, if the inner binary is unresolved, then this triple system would be considered as a binary during the counting; if the inner binary is resolved, the system would be considered as one having multiple comoving companions and hence are excluded in our counting. Nonetheless, the contribution of marginally resolved hierarchical triples (those only resolved at small distances) is expected to be small and should not affect our conclusions.

With the absolute magnitude criterion for the companions, we end up with 7,671 (7,266) comoving pairs with separations of $10^3$-$10^4$\,AU for the LASP (DD-Payne) sample. Among them, there are 330 pairs (660 LAMOST stars) where both stars in the pair were observed by LAMOST. Some of these pairs have been studied to show that the components of wide binaries have similar metallicity and elemental abundances \citep{Andrews2018, Andrews2019}. Since the definition of our wide-binary fraction is essentially the probability that a randomly selected star is in a wide binary system, the proper statistics requires that we account for both LAMOST stars in $N_{\rm LAMOST}$ and $N_{\rm companion}$ even if they belong to the same pair. LAMOST targeting does not depend on the binarity of stars \citep{Carlin2012}, and therefore no direct systematics is inherited from the targeting. 


\begin{figure*}
	\centering
	\includegraphics[height=.34\linewidth]{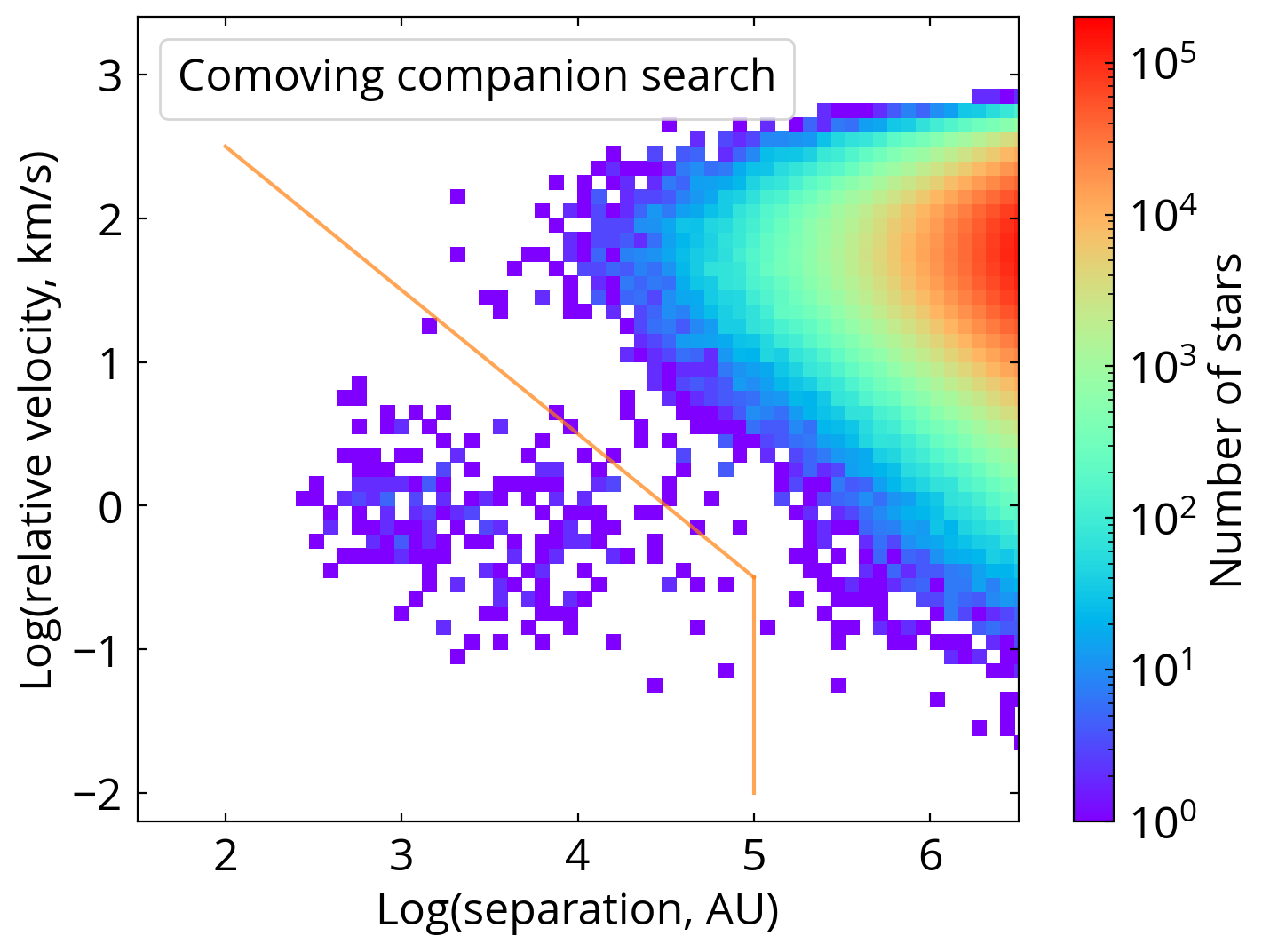}
	\includegraphics[height=.34\linewidth]{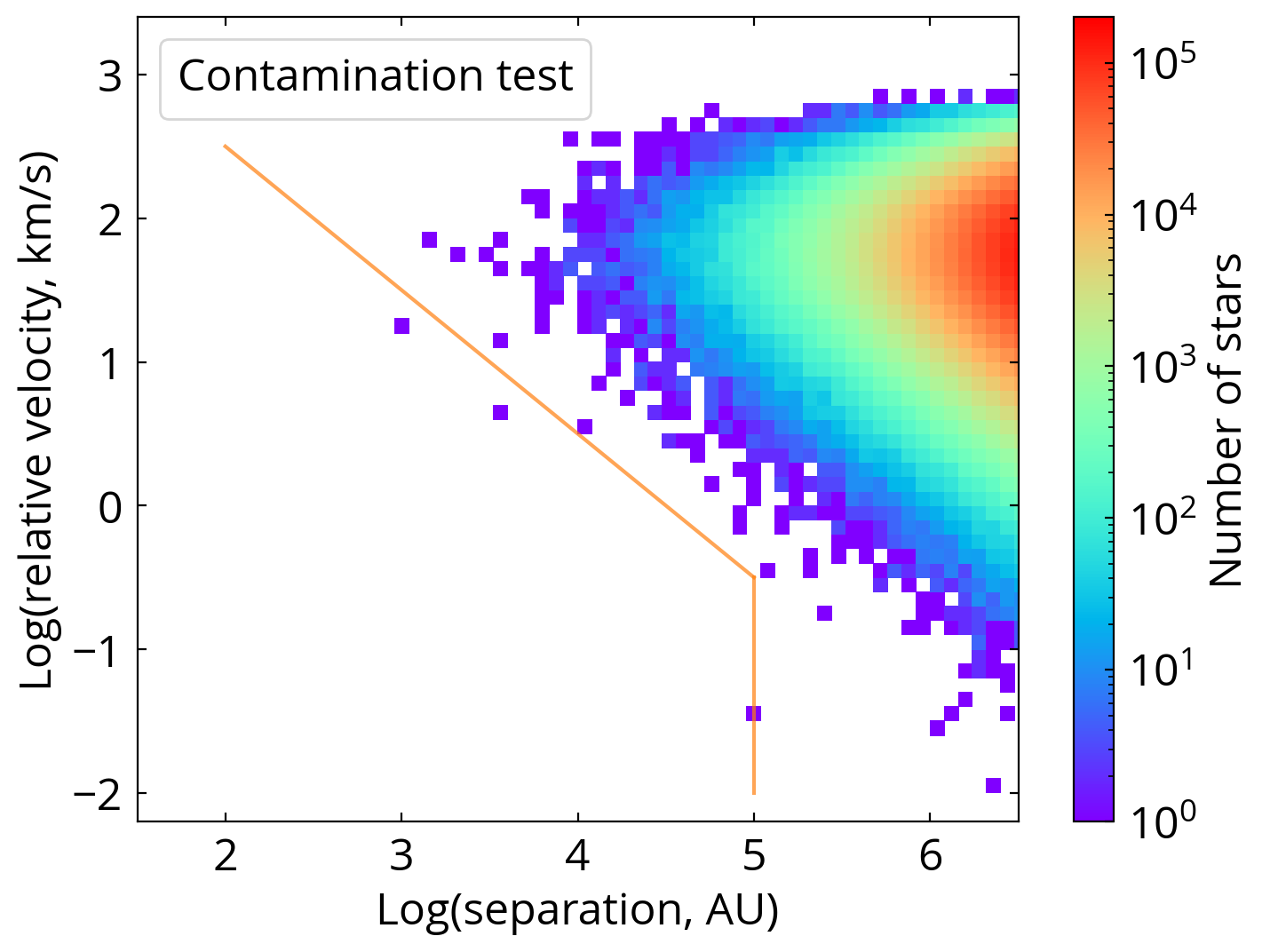}
	\caption{The comoving companion search (left) and the contamination test (right) for 4000 randomly selected LAMOST stars, where 2000 of them have [Fe/H]$<-0.75$ and the other 2000 have [Fe/H]$>0$. The contamination test flips the sign of the Galactic latitude and the proper motion in the direction of Galactic latitude; therefore, all pairs in the right panel are chance projection. The orange solid line is the empirical selection for comoving companions. The level of contamination from chance projection plays a negligible role in our results. }
	\label{fig:contamination-test}
\end{figure*}

\subsection{Contamination test}

We test the contamination of our comoving search by flipping the sign of the Galactic latitude and the proper motion in the direction of Galactic latitude of the LAMOST stars. The comoving search for a given LAMOST star only considers the Gaia sources nearby its flipped location and does not include other flipped LAMOST stars. Thus, the flipped LAMOST stars have similar surrounding stellar densities as their original locations, but now all nearby stars are chance projection. We randomly select 2000 LAMOST stars with [Fe/H]$<-0.75$ and 2000 with [Fe/H]$>0$ to investigate if the contamination level depends on the metallicity. We ensure that the sky regions of the flipped LAMOST stars are covered by Gaia DR2 with visibility periods $>8$. The solar motion and the Galactic disk differential rotation contribute different proper motions depending on the sky location, which need to be taken into account in the contamination test. We remove their contributions to proper motions using the local shear approximation \citep{Olling2003} with the solar motion from \cite{Schonrich2010} and the Oort constants from \cite{Bovy2017}. Therefore, the solar motion and the disk differential rotation do not contribute the relative velocity in the contamination test.

Fig.~\ref{fig:contamination-test} shows the comoving search result of the 4000 LAMOST stars (left) and their contamination test (right). The orange line is the empirical demarcation line designed to have a similar slope as the chance projection \citep{Hwang2020c}, and we only consider wide binaries at separations between $10^3$ and $10^4$\,AU in this paper. Among these 4000 LAMOST stars, 105 of them have wide companions in the left panel (with the absolute magnitude cut on the companions), and only one chance projection in the right panel (there are other two chance-projection pairs below the demarcation line, but their separations are not between $10^3$ and $10^4$\,AU). Therefore, the contamination level is about two order-of-magnitude smaller than the wide-binary fraction regardless of the metallicity. This contamination is lower than in \cite{Hwang2020c} (which is purely based on Gaia) because most of the LAMOST stars are located in lines-of-sight towards the outer disc \citep{Zhao2012}. In short, the level of contamination from chance alignments plays a negligible role in our results.

The astrometric measurements may be affected by the presence of subsystems. For example, the orbital motion \citep{Belokurov2020} and the photometric variability \citep{Hwang2020a} of the unresolved systems may induce astrometric noise. The presence of a marginally resolved source (angular separations of a few $\times0.1$\,arcsec) also downgrades the astrometric measurement quality because of the non-point-spread-function light profile \citep{Hwang2020a}. The blending of unresolved spectra may result in unreliable or flagged metallicity. These possibilities would reduce the completeness of the wide binaries that have subsystems; however, they are unlikely to affect our results significantly. First, the parameter space for (marginally) unresolved systems to have corrupted astrometric measurements due to the orbital motions or photometric variability is narrow, especially that the angular separation of such system needs to be large and the orbital or photometric timescale needs to be comparable or shorter than Gaia's temporal baseline. Second, these possibilities affect parallaxes more than the proper motions because most of our sample have parallaxes close to 2\,mas, while their median total proper motions is about 20\,mas\,yr$^{-1}$. This is the reason we use a more relaxed parallax criterion in the comoving companion search (either parallax difference $< 0.2$\,mas or the line-of-sight distance difference $<20$\,pc). A more relaxed parallax criterion may result in a higher contamination, but Fig.~\ref{fig:contamination-test} shows that the contamination level remains negligible. Third, if our results are due to the presence of subsystems, then we would expect our results to change for a sample at different distances and for wide binaries with different separations. In the Appendix~\ref{sec:appendix-test}, we show that our conclusions remain unchanged when different selection criteria are used.


\subsection{Distinguishing thin disk, thick disk, and halo stars}
\label{sec:distinguish-pop}

\begin{figure}
	\centering
	\includegraphics[width=\linewidth]{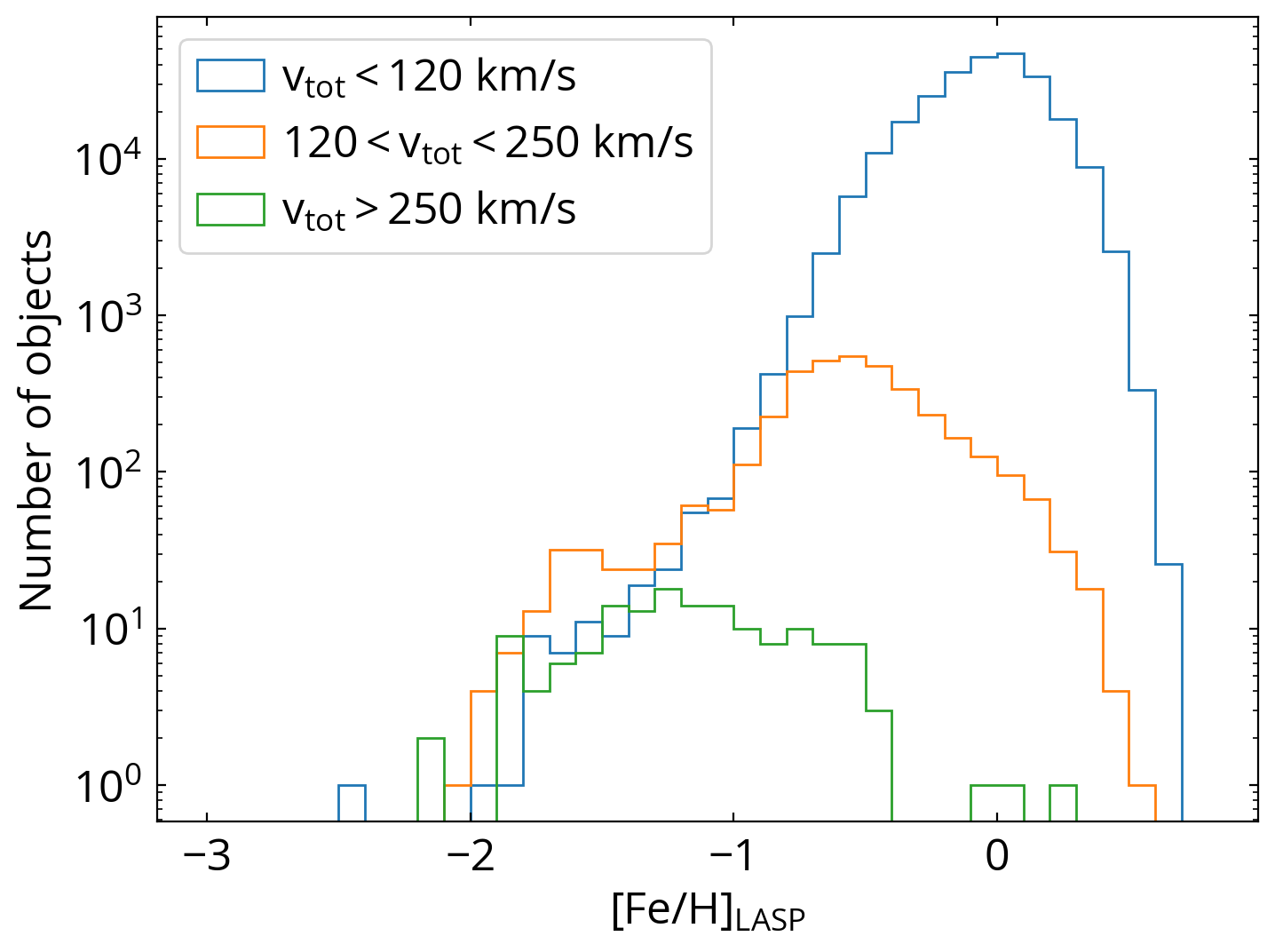}
	\includegraphics[width=\linewidth]{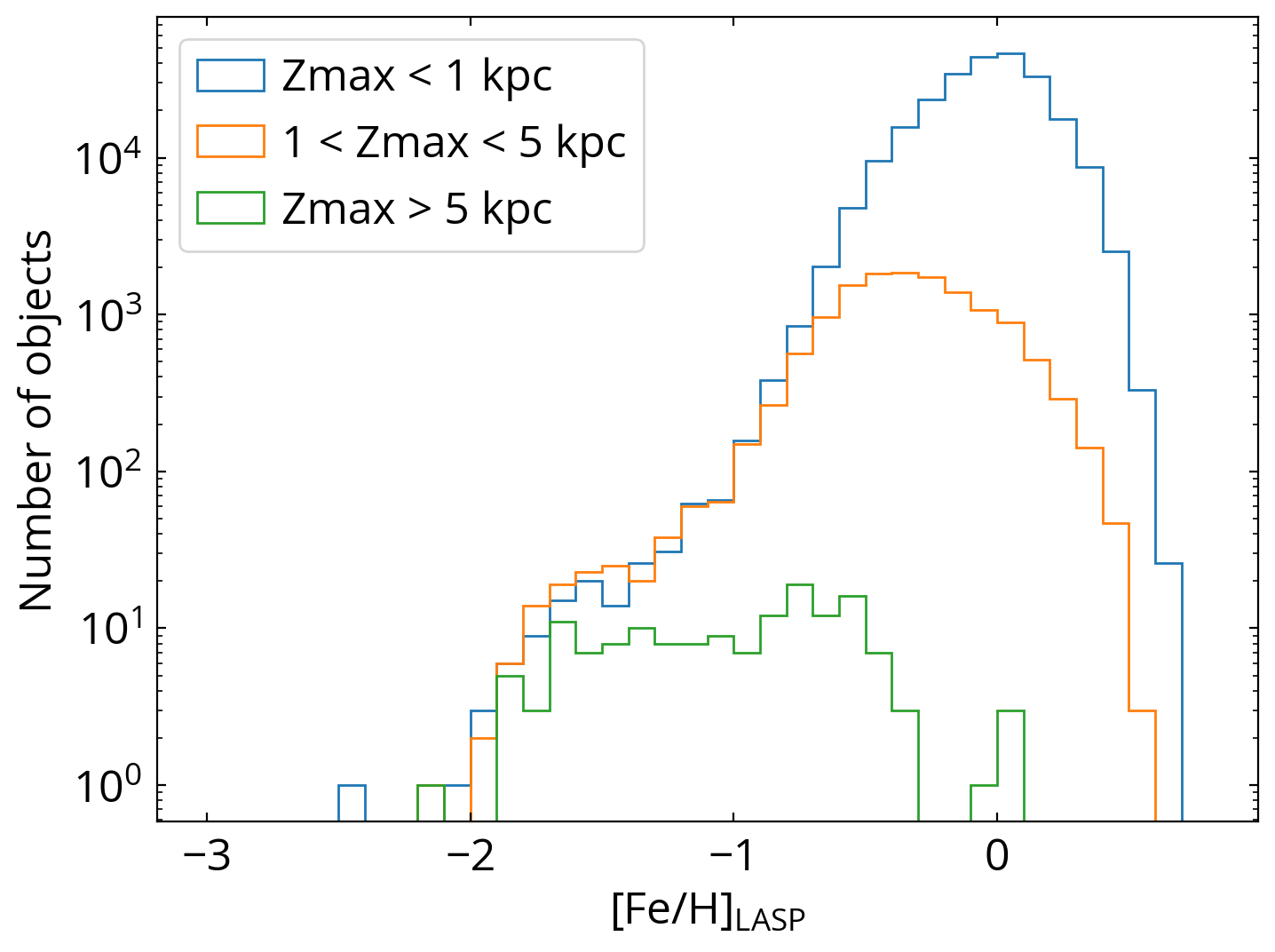}
	\caption{The LASP metallicity distribution for the thin-disk, thick-disk, and halo stars, selected using total velocity (top) and the maximum Galactic height (bottom). }
	\label{fig:metallicity-dist}
\end{figure}

We consider two methods to distinguish the thin disk, thick disk, and halo stars: (1) the maximum Galactic height of the Galactic orbits (maximum vertical excursion, $z_{max}$); and (2) total 3-dimensional velocity ($v_{tot}$), computed from the projected velocity from Gaia and the radial velocity from LAMOST LASP, with respect to the local standard of rest \citep{Schonrich2010}. We use \texttt{galpy}\footnote{http://github.com/jobovy/galpy} \citep{Bovy2015} to derive the $z_{max}$ of the Galactic orbits. Specifically, we use the fast estimation of orbit parameters via the St\"ackel approximation, and the estimation for $z_{max}$ is precise to a level better than $\sim1$\% \citep{Mackereth2018}. We use the Milky Way potential \texttt{MWPotential2014} from \cite{Bovy2015} and a solar motion with respect to the local standard of rest from \cite{Schonrich2010}. 

We use the Gaia DR2 mock catalogue \citep{Rybizki2018} to test our selection for thin-disk, thick-disk, and halo stars. The Gaia DR2 mock catalogue is generated using \texttt{Galaxia} \citep{Sharma2011} that samples stars from the Besan\c con Galactic model \citep{Robin2003}. To match the properties of our LAMOST F- and G dwarfs, we select main-sequence mock stars with $0.6<$BP-RP$<1.1$ and parallax $>2$\,mas. Following \cite{Hwang2020b}, we assign weights to the mock stars so that they have a similar sky distribution as our LAMOST sample. Then we use \texttt{galpy} to derive the $z_{max}$ for the mock stars.

Using Gaia DR2 mock catalogue, we find that 91\% of $v_{tot}<120$\kms\ stars belong to the thin disk, 87\% of $120<v_{tot}<250$\kms\ stars belong to the thick disk, and 88\% of $v_{tot}>250$\kms\ stars belong to the halo. For the $z_{max}$ selection,  92\% of $z_{max}<1$\,kpc stars are thin disk, 80\% of $1<z_{max}<5$\,kpc are thick disk, and 35\% of $z_{max}>5$\,kpc are halo stars. Therefore, we consider $v_{tot}$ as a better selection for the halo sample than $z_{max}$. Their metallicity distributions are shown in Fig.~\ref{fig:metallicity-dist}. The low-metallicity tail at [Fe/H]$<-1$ in the thin-disk stars may be partially contributed by the contamination from the thick-disk stars. By using the $v_{tot}$ ($z_{max}$) selection, we have 7602 (7335), 67 (334), and 2 (2) wide binaries in the thin disk, thick disk, and halo respectively. We caution readers for the results for the halo in this study due to its small sample, and one of the $z_{max}$-selected halo wide binaries has [Fe/H]=$-0.46$ and is likely a thick disk contaminant.

\section{The metallicity and age dependence of the wide-binary fraction}
\label{sec:result}

\begin{figure*}
	\centering
	\includegraphics[height=.45\linewidth]{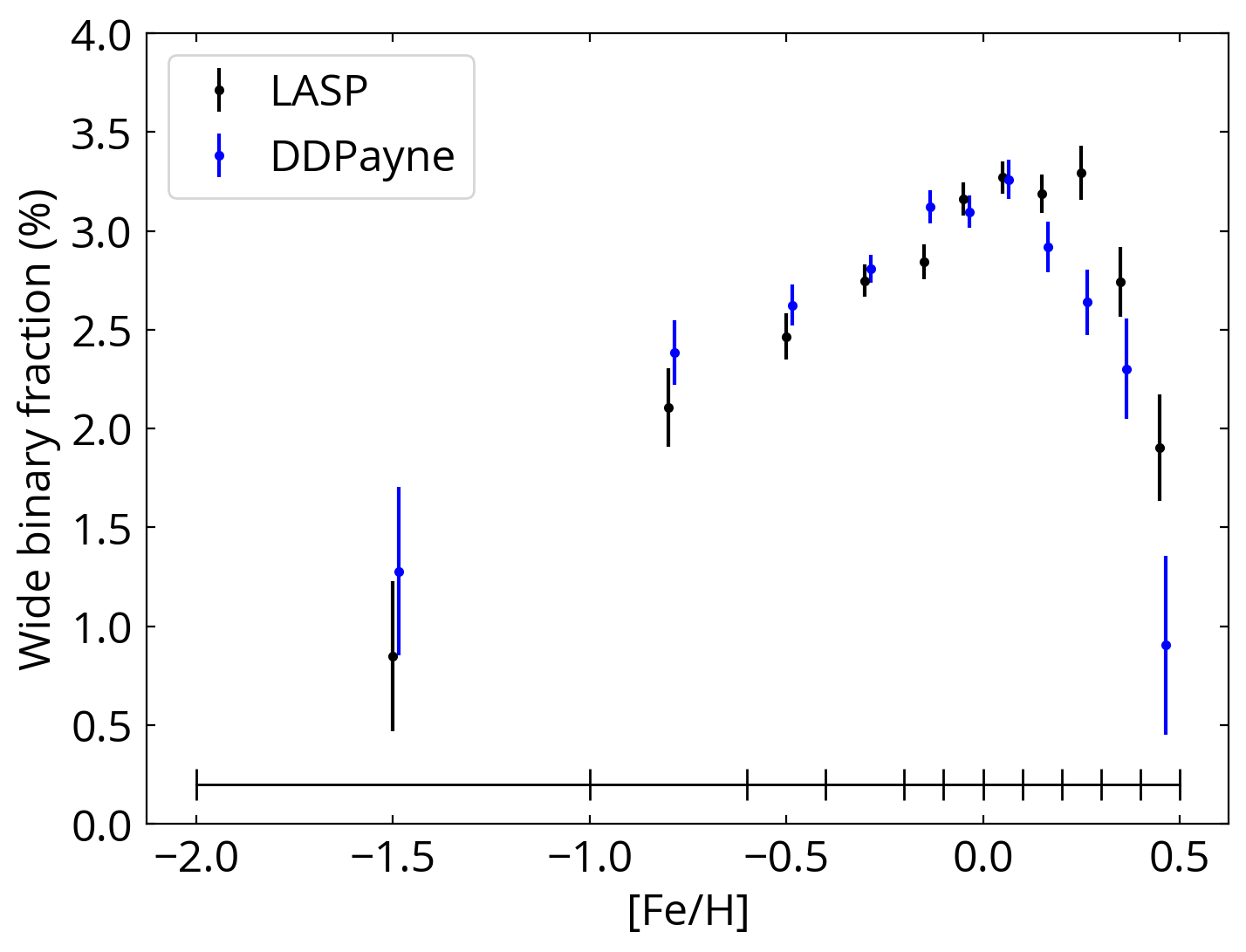}
	
	\caption{The metallicity dependence of the wide-binary fraction ($10^3$-$10^4$\,AU). The black points use the [Fe/H] derived from LAMOST Stellar Parameter pipeline (LASP), and the blue points use the LAMOST [Fe/H] measured by DD-Payne. The black and the blue points are slightly offset along the horizontal axis for clarity. The ticks at the bottom show the bin size, and the markers are located at the center of the bin. Both results show that, as [Fe/H] increases, the wide-binary fraction first increases at low [Fe/H], peaks at [Fe/H]$\simeq 0$, and then decreases at high [Fe/H]. }
	\label{fig:metallicity}
\end{figure*}

\begin{figure*}
	\centering
	\includegraphics[height=.34\linewidth]{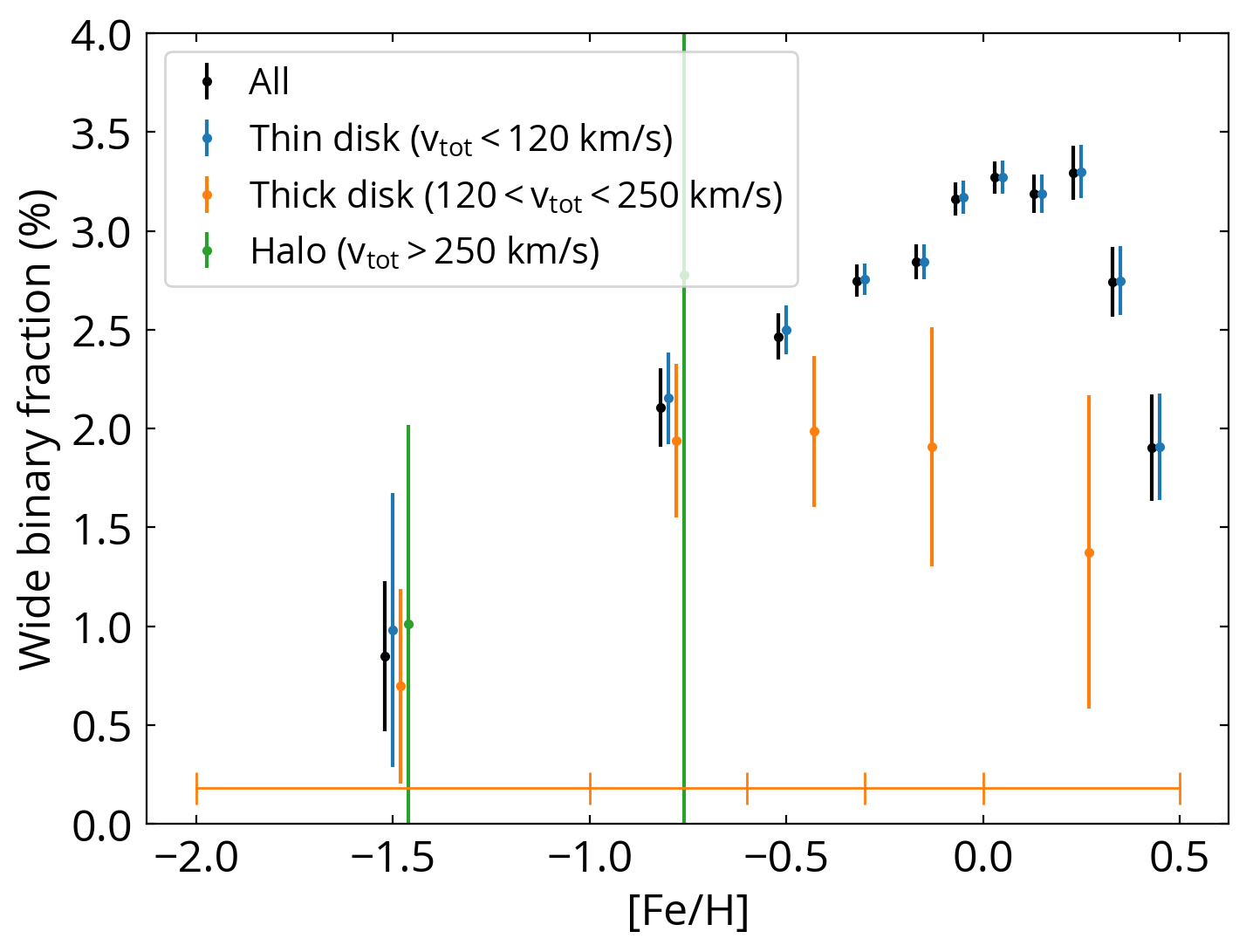}
	\includegraphics[height=.34\linewidth]{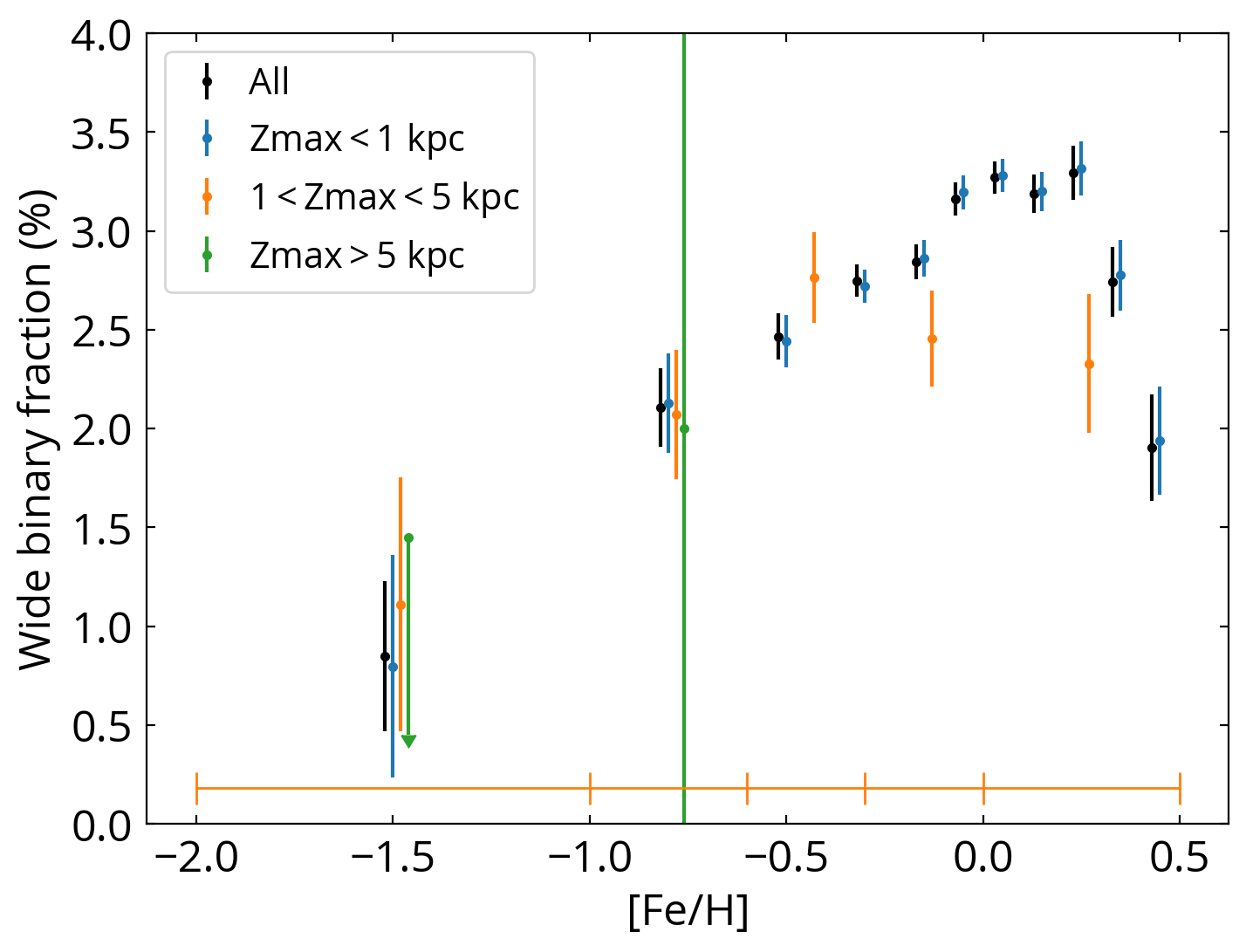}
	\caption{The metallicity dependence of the wide-binary fraction in the thin disk, thick disk, and halo. The LASP metallicity is adopted here, and results from DD-Payne are similar. For comparison, the black points show the same result from Fig.~\ref{fig:metallicity}. The bin size of the thin-disk sample is the same as Fig.~\ref{fig:metallicity}, and that of the thick-disk and halo samples is shown as the ticked orange line at the bottom. The left panel uses the total velocity to select different populations, and the right panel uses the maximum Galactic height of the orbits. The results of these two selections are in good agreement with each other. The metallicity dependence of wide binaries is dominated by the thin-disk stars. The wide-binary fraction of the thick disk follows a similar trend as the thin disk at low [Fe/H], and then become flat with increasing metallicity at [Fe/H]$>-0.5$. The wide-binary fraction in the halo is not well constrained due to small number statistics.}
	\label{fig:metallicity-population}
\end{figure*}

\begin{figure}
	\centering
	\includegraphics[width=\linewidth]{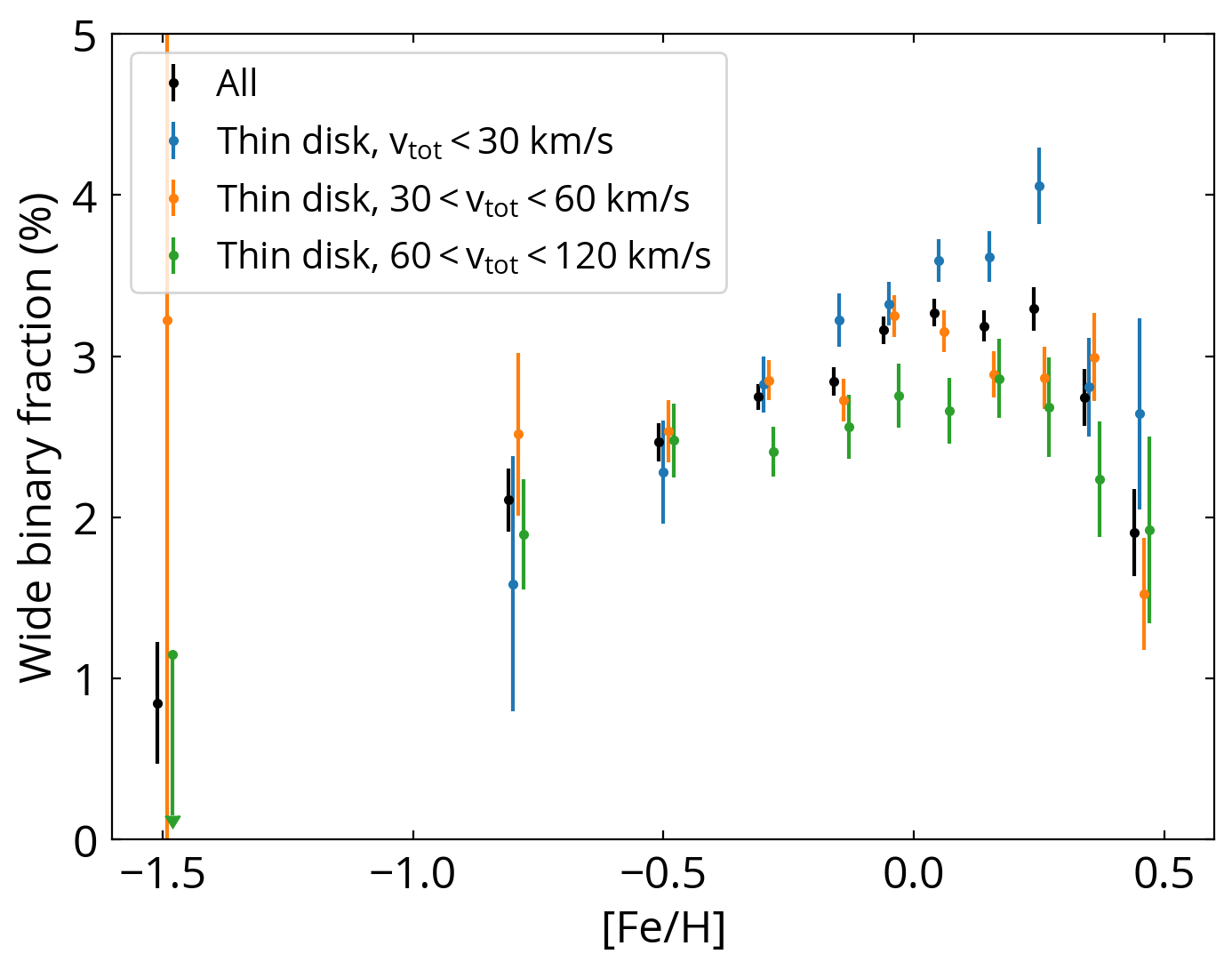}
	
	\caption{The wide-binary fraction as a function of metallicity for the thin-disk stars in bins of total velocity. Points are slightly offset horizontally for clarity. For comparison, the black points show the result from all stars in Fig.~\ref{fig:metallicity}. The velocity is a proxy of the stellar age, where older stars typically exhibit larger velocities. The wide-binary fraction of the low-velocity (young) stars has a stronger metallicity dependence. }
	\label{fig:metallicity-v}
\end{figure}

\begin{figure}
	\centering
	\includegraphics[width=\linewidth]{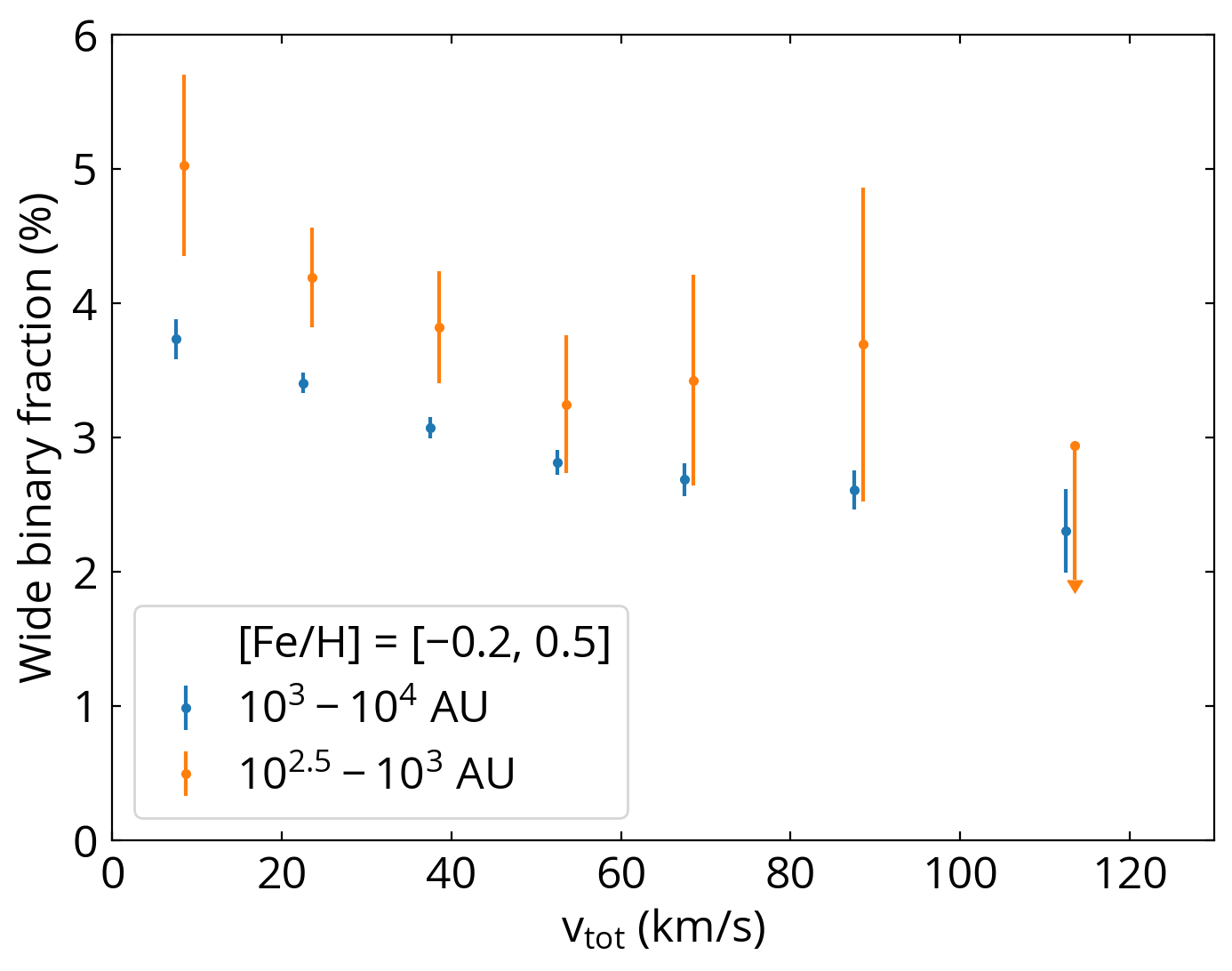}
	
	\caption{The relation between the wide-binary fraction and the total velocity, where the total velocity is a proxy of stellar age. Here we only consider [Fe/H] between $-0.2$ and $+0.5$. The wide-binary fraction shows a prominent decrement with increasing velocity (and hence increasing stellar age) at $v_{tot}<50$\kms. This age dependence is also present in the wide binaries with smaller separations of $10^{2.5}$\,AU. }
	\label{fig:metallicity-vtot}
\end{figure}

Fig.~\ref{fig:metallicity} shows the wide-binary fraction as a function of stellar metallicity. The black points use the LASP metallicity, and the blue points use the metallicity derived by DD-Payne. The metallicity bins span from [Fe/H]$=-2$ to $+0.5$ and are indicated by the ticked line at the bottom of Fig.~\ref{fig:metallicity}, with the markers located at the center of each metallicity bin. The bin sizes and the numerical values are available in Appendix~\ref{sec:appendix-data}. The black and the blue points are slightly offset horizontally for clarity. Error bars of the wide-binary fractions are Poisson uncertainties.

The overall metallicity dependence is similar for LASP metallicity and DD-Payne metallicity: the wide-binary fraction first increases with increasing metallicity, peaks at [Fe/H]$\simeq 0$, and then decreases at the high metallicity end. The metallicity where the wide-binary fraction peaks is slightly different between LASP and DD-Payne, which is likely due to the systematic metallicity offset of $0.07$\,dex between LASP and DD-Payne (Sec.~\ref{sec:LAMOST}). Otherwise, our result is robust over different metallicity pipelines. We focus on the results using LASP metallicity for the rest of the figures, and we do not find significant difference from those using DD-Payne metallicity.

We perform the Kolmogorov-Smirnov test to quantify the significance of the difference in the metallicity distributions between the stars with wide companions and the entire main sample (with criteria described in Sec.~\ref{sec:main-selection}). The $p$-value, the probability that two metallicity distributions are sampled from the same parent distribution, is $2\times10^{-10}$. Therefore, the difference is statistically significant, suggesting that wide binaries show robust metallicity dependence and are not a randomly drawn subsample of the parent distribution.

Since different populations may dominate at different metallicities, we further divide the sample into thin-disk, thick-disk, and halo stars using the total velocity (left panel) and the $z_{max}$ (right panel) in Fig.~\ref{fig:metallicity-population}. For comparison, the black points are the LASP points from Fig.~\ref{fig:metallicity}. For the thin-disk samples, we adopt the same metallicity bins as for the full sample (black points), and we use larger metallicity bins (the ticked orange line at the bottom of Fig.~\ref{fig:metallicity-population}) for the thick-disk and halo samples to reduce the Poisson uncertainties. Overall, the velocity-selected samples are in agreement with the $z_{max}$-selected samples. The thin-disk sample shows a similar trend as in Fig.~\ref{fig:metallicity}, meaning that the global metallicity dependence is dominated by the thin disk stars, which constitutes a large fraction of the LAMOST sample. The thick-disk sample follows the metallicity relation of the thin-disk stars at [Fe/H]$<-0.4$. At [Fe/H]$>-0.4$, the wide-binary fraction of the thick-disk sample is much lower than that of the thin disk. The halo sample has a wide-binary fraction of $\lesssim 1.5$\% in the metallicity bin of $-2<$[Fe/H]$<-1$. While it is consistent with the wide-binary fractions in the thin-disk and thick-disk stars at the same metallicity, the wide-binary fraction in the halo is not well constrained given that there are only two wide binaries in the halo sample.

Directly age-dating main-sequence stars is difficult, and in most cases, impossible. However, kinematics of main-sequence stars nonetheless gives a \textit{statistical} estimate of the ages, especially for the thin-disk stars because their dynamical evolution is mostly secular \citep{Dehnen1998,Nordstrom2004,Reid2009,Sharma2014a, Ting2019a}. Hence, in Fig.~\ref{fig:metallicity-v}, we use total velocities to investigate the stellar age dependence of the wide-binary fraction in the thin disk as a function of metallicity. Here we select thin-disk stars by $z_{max}<1$\,kpc and bin the sample into the low-velocity ($v_{tot}<30$\kms), the middle-velocity ($30<v_{tot}<60$\kms), and the high-velocity ($60<v_{tot}<120$\kms) sample.

Fig.~\ref{fig:metallicity-v} shows that wide-binary fractions of all velocity sub-samples have a similar metallicity trend, but such metallicity dependence is strongest in the low-velocity, young sample. In the metal-poor regime ([Fe/H]$\lesssim-0.5$), the wide-binary fractions of stars are about the same, irrespective of their velocity. In contrast, the lowest-velocity sample has a higher wide-binary fraction at [Fe/H]$\sim0$ than that of the higher-velocity samples. This result also means that the metallicity dependence of the wide-binary fraction in the thin disk is not due to the varying levels of contamination from the thick-disk stars with respect to metallicity, in which case we would expect a weaker metallicity dependence in the low-velocity thin-disk sample where the contamination is the lowest. 

Fig.~\ref{fig:metallicity-v} demonstrates that younger stars have a higher wide-binary fraction, especially at [Fe/H]$\sim0$. However, we caution that Fig.~\ref{fig:metallicity-v} does not necessarily mean that there is a metallicity-dependent age evolution for wide binaries, because each metallicity bin may have different age distributions. It is possible that the lack of age evolution in the metal-poor regime ([Fe/H]$<-0.5$) is simply because these metallicity bins lack young stars (e.g. \citealt{Casagrande2011,Lin2020}).

To explore the stellar age dependence further, in Fig.~\ref{fig:metallicity-vtot} we investigate the wide-binary fraction as a function of total velocity. We also present the wide-binary fraction for separations between $10^{2.5}$ and $10^3$\,AU, where we apply a parallax cut $>6.3$\,mas for the sample so that $10^{2.5}$\,AU corresponds to the angular resolution of 2\,arcsec. Furthermore, we adopt an absolute G-band magnitude criterion of 12.5\,mag for the $10^{2.5}$-$10^3$\,AU case. The result is similar but noisier if we use the original criterion of 10\,mag. Here we only consider metallicity between $-0.2$ and $0.5$ because they cover a wider age distribution compared to the metal-poor stars. In case that massive companions may induce additional age dependence, we test the selection by requiring that the companions be fainter than the LAMOST stars, and the result remains nearly the same.

Fig.~\ref{fig:metallicity-vtot} shows that stars having $v_{tot}<50$\kms\ have a higher wide-binary fraction with separations down to $10^{2.5}$\,AU. Based on the Gaia DR2 mock catalogue \citep{Rybizki2018} which sample mock stars from the Besan\c con Galactic model \citep{Robin2003}, the total velocity of $\sim50$\kms\ corresponds to a mean stellar age of $\sim 5$\,Gyr. Therefore, the wide-binary fraction seems to be higher in younger stars with ages $\lesssim $ a few Gyr.

\section{Discussion}

\label{sec:discussion}

We have found that in the thin disk, the wide-binary fraction increases with metallicity at [Fe/H]$\lesssim0$, and then decreases with metallicity in the super-solar regime. Furthermore, based on the kinematics, the enhanced wide-binary fraction at [Fe/H]$\simeq 0$ is age dependent, with a higher wide-binary fraction in younger stars. In the following sections, we compare these results to those from past studies, and seek an interpretation that would simultaneously explain the metallicity and age dependence of the wide-binary fraction.

\subsection{Comparison with previous work}

In the pre-Gaia era, some studies concluded that the wide-binary fraction was relatively independent of metallicity \citep{ZapateroOsorio2004,Zinnecker2004}, while some found a lower wide-binary fraction for metal-poor stars \citep{Rastegaev2008,Jao2009, Lodieu2009, Zhang2013, Ziegler2015}. \cite{Moe2019} argued that a lower wide-binary fraction of the metal-poor stars from high-resolution imaging studies may be a selection effect, because photometric selection of metal-poor stars may exclude unresolved metal-poor binaries since they are brighter than the metal-poor single stars and may be considered as metal-rich single stars. Our results do not involve any photometric estimates of metallicity and are free from such selection effect.

In the Gaia era, with proper motions and parallaxes available for billions of stars, a large sample of comoving pairs has been made possible \citep{Oh2017, El-Badry2018b,Jimenez-Esteban2019}. In particular, \cite{El-Badry2019a} study the metallicity dependence of wide binaries with separations from $50$ to $50,000$\,AU by combining the comoving pair sample from Gaia DR2 and wide-field spectroscopic surveys. For binaries with separations $\gtrsim 250$\,AU, they conclude that the binary fraction remains constant with respect to metallicity.

Our sample bears some similarities to the one from \cite{El-Badry2019a}, but here we complement the study by expanding the sample to 500\,pc. Their sample is restricted within 200\,pc. Therefore, we have a larger sample at larger distances, which strongly improves the constraints on the thick-disk and halo stars at the low-metallicity end. Our sample enables us to further dissect the wide-binary fraction as a function of metallicity and age, while \cite{El-Badry2019a} do not take the kinematics and ages into account.

While our findings of the strong metallicity dependence for the wide-binary fraction seem at odds with their conclusion, \cite{El-Badry2019a} do comment that there is a slight excess of wide binaries at [Fe/H]$\simeq 0$, consistent with our results. They suspect that such excess may be due to the age effect such that old wide binaries are disrupted by gravitational perturbations from other stars and molecular clouds. In the following section, we investigate this possibility in detail, and will argue that gravitational perturbations are unlikely to play a dominant role.

\subsection{Wide binary disruption}
\label{sec:disruption}

When time passes, wide binaries may be disrupted by passing stars, molecular clouds, and Galactic tidal fields \citep{Bahcall1985,Weinberg1987,Chaname2004, Yoo2004,Quinn2009,Jiang2010a}. Binaries with wider separations are easier to be disrupted due to the weaker binding energy. In particular, given the stellar density in the solar neighborhood, theoretical estimates show that binaries with separations $> 0.1$\,pc ($2\times10^4$\,AU) would be disrupted within 10\,Gyr \citep{Weinberg1987}. Therefore, fewer binaries with separations $>10^4$\,AU are expected in the old disk stars \citep{Bahcall1981,Retterer1982,Weinberg1987}. \cite{Tian2020} may detect this effect in their `halo sample' selected by the high tangential velocities ($>85$\kms), and the authors argue that their results cannot be explained by the binary disruption due to the low density in the halo. While this hints that there might be other effects beyond gravitational perturbations which shape the wide-binary fraction, we note that their results might not be conclusive, as a tangential velocity cut at $>85$\kms\ likely results in predominantly old thin-disk stars and thick-disk stars, instead of halo stars.

While the disruption of binaries by the gravitational perturbations (passing stars, molecular clouds, and Galactic tidal fields) may be able to make the wide-binary fraction lower in the metal-poor stars because they are on average older, this explanation alone is at odds with some results presented in this study. First, theoretical arguments have shown that the disruption lifetime of $10^3$\,AU binaries is $\sim100$\,Gyr, much longer than the age of Universe \citep{Weinberg1987}. Furthermore, if binary disruption were to play an important role, we expect wider binaries should be preferentially disrupted. However, our data do not show a significant difference in the age evolution between $10^{2.5-3.0}$\,AU and $10^{3-4}$\,AU binaries (Fig.~\ref{fig:metallicity-vtot}). Also, binary disruption is not able to explain the anti-correlation between wide-binary fraction and metallicity at [Fe/H]$>0$. Therefore, we conclude that the age and metallicity dependence of the wide-binary fraction cannot solely be explained by binary disruption. 


\subsection{Wide binary formation and evolution}

Since wide binary disruption cannot be the whole story, here we investigate whether the metallicity and age dependence arise from wide binary formation. Wide binaries with separations of $10^3$-$10^4$\,AU can be formed through multiple channels, including the turbulent core fragmentation \citep{Padoan2002,Fisher2004,Offner2010}, dynamical unfolding of unstable compact triples \citep{Reipurth2012,Elliott2016}, the dissolution of star clusters \citep{Kouwenhoven2010, Moeckel2011}, and the pairing of adjacent pre-stellar cores \citep{Tokovinin2017}. However, not all of these channels can provide the observed metallicity and age dependence of the wide-binary fraction.

\subsubsection{The negative metallicity dependence}

We first tackle the decrease of the wide-binary fraction with metallicity at [Fe/H]$>0$. The dynamical unfolding of compact triples may be able explain such metallicity dependence. In this scenario, triple stars are born in compact, unstable configurations, and then they evolve to a hierarchical architecture with one companion scattered into a wide orbit \citep{Reipurth2012}. As a result, the formation of wide binaries is influenced by the occurrence of close binaries, so the metallicity dependence of the wide binaries is inherited from the formation of compact systems through disk and other small-scale fragmentation. If wide companions were preferentially formed via this scenario, then the wide-binary fraction would follow a similar metallicity dependence as close binaries. Indeed, close binaries also show a declining occurrence rate as a function of metallicity \citep{Grether2007, Raghavan2010,Yuan2015,Badenes2018,Moe2019,El-Badry2019a,Mazzola2020}, as is observed for the super-solar metallicity sample in this study. This may be in line with the excess of equal-mass binaries (`twin' binaries) at separations $>1000$\,AU, which also suggests that these wide binary twins are formed at close separations initially ($a<100$\,AU) and then their orbits are widened by the dynamical interaction with the birth environments \citep{El-Badry2019}.

The connection between wide companions and close binaries is supported by other observational studies. For instance, 96\% of close binaries with orbital periods $<3$\,days have tertiary companions \citep{Pribulla2006,Tokovinin2006}. \cite{Hwang2020c} find that the occurrence rate of the wide companions at $10^3$-$10^4$\,AU around main-sequence contact binaries is a factor of $3$ higher compared to that of the field stars. Conversely, about half of wide pairs with separations of $10^3$-$10^4$\,AU are hierarchical multiples \citep{Raghavan2010,Moe2017,Moe2019b}. 

The enhanced occurrence rate of tertiary companions around close binaries possibly suggests that tertiary companions play a critical role in the orbital migration of the inner binary through the Kozai-Lidov mechanism, where the outer tertiary companion excites the high eccentricity of the inner binary \citep{Kozai1962, Lidov1962, Kiseleva1998,Eggleton2001,Eggleton2006,Fabrycky2007,Naoz2013,Borkovits2016}. Nonetheless, noting that the Kozai-Lidov mechanism is only effective under certain inner-to-outer separation ratios and mutual inclinations, it remains unclear whether this mechanism can be responsible for the majority of those triple systems consisting of close binaries \citep{Moe2018, Hwang2020b,Hwang2020c}. Alternatively, the enhanced occurrence rate of tertiary companions may be a result of compact multiple stars forming from disk fragmentation \citep{Tokovinin2020}, then the occurrence of these wide companions would follow the similar anti-correlation with metallicity as the close binaries, in line with the scenario of dynamical unfolding of compact triples.


To sum up, we argue that the negative metallicity dependence of the wide-binary fraction is inherited from that of the close-binary fraction through the dynamical unfolding of triple stars. Nonetheless, the metallicity dependence of the wide-binary fraction is clearly non-monotonic. It raises a question why this anti-correlation is only present at [Fe/H]$>0$, while that of the close-binary fraction spans from [Fe/H]$=-3$ to $+0.5$. This implies that there is another limiting factor dominating the wide binary formation at the metal-poor regime, which we investigate in detail in the next section.

\subsubsection{The positive metallicity dependence }

During the pre-main sequence phase (ages $<$ a few Myr), wide binaries can form through the turbulent core fragmentation and the random pairing of adjacent pre-stellar cores. While the wide-binary fractions from these two mechanisms are not explicitly dependent on metallicity, as is shown in the hydrodynamical simulations \citep{Bate2005, Bate2014,Bate2019}, wide binaries themselves are sensitive to the formation environments. In particular, most if not all stars form in clustered environments \citep{Lada2003}, and about 20-30\% of stars originate from bound clusters \citep{Bressert2010,Kruijssen2012,Chandar2017}. Environments with a higher stellar density have small stellar separations, making wide binaries more difficult to survive. Furthermore, the higher velocity dispersion accompanied by the higher stellar density makes the random pairing less likely. Indeed, observational studies have found that wide-binary fractions are higher in the low-density star-formation regions compared to the higher-density clustered environment \citep{Simon1997,Kraus2009,Tobin2016, Elliott2016, Joncour2017, Deacon2020}. Therefore, density of the formation environment plays a critical role in the wide binary formation (e.g. \citealt{Marks2011a, Marks2011, Marks2012}).

When the gas is removed after $\sim10$\,Myr \citep{Bastian2005,Fall2005,Mengel2005}, the cluster expands in response to the change in the gravitational potential \citep{Goodwin2006,Goodwin2009}. At this cluster dissolution phase, two unbound stars that are originally close in the phase space may pair together and become a wide binary \citep{Kouwenhoven2010,Moeckel2011}. Using Monte Carlo and $N$-body simulations, \cite{Kouwenhoven2010} further find that the wide-binary fraction decreases strongly with increasing cluster mass, where the main driving factor may be associated with the increasing velocity dispersion that makes two stars less likely to pair in the phase space.

In the earlier Universe, star formation environments tend to have a higher pressure and density than the present day, and high-mass clusters are preferentially formed in such environments \citep{Harris1994,Elmegreen1997,Kravtsov2005,Kruijssen2014, Ma2020}. A higher-density environment reduces the wide binary formation from the turbulent core fragmentation and the random pairing of adjacent pre-stellar cores, and also fewer wide binaries can form from the dissolution of higher-mass clusters. As a result, the wide-binary fraction would be lower in the older stars, which explains the age dependence in Fig.~\ref{fig:metallicity-v} and Fig.~\ref{fig:metallicity-vtot}. Furthermore, because metal-poor stars are on average older stars, this naturally explains the positive correlation between the wide-binary fraction and metallicity.

\subsection{A holistic view and future outlook}
\label{sec:holistic}

\begin{figure*}
	\centering
	\includegraphics[height=0.45\linewidth]{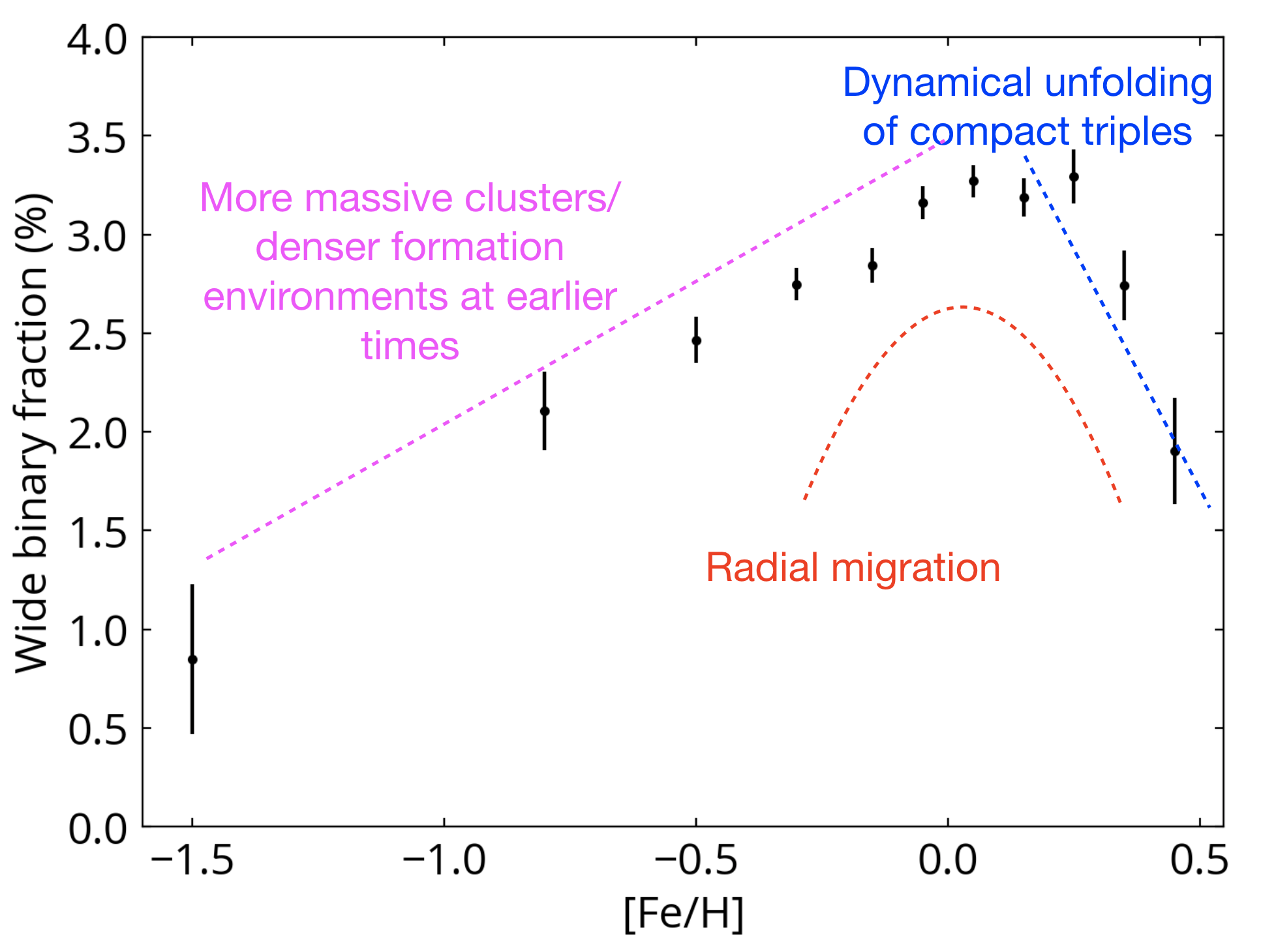}
	
	\caption{Schematic illustration of the metallicity dependence of various proposed wide binary formation channels in this study. The observed metallicity dependence (black points) is likely the consequence of multiple formation channels of wide binaries. The dashed lines show the metallicity trend of the proposed mechanisms, and their values and slopes are only for schematic illustration. The dynamical unfolding of compact triples (blue line) follows the metallicity anti-correlation of the close-binary fraction. The trend at [Fe/H]$<0$ can be due to that the density of the formation environments and the cluster mass are higher at earlier times (pink line). The environmental effect ceases to play a role at the high metallicity because the formation time is similar at [Fe/H]$>0$, and therefore the anti-correlation from the dynamical unfolding of triple stars manifests itself at high metallicities. Radial migration may also play a role in enhancing the wide-binary fraction around the solar metallicity (red line). }
	\label{fig:explanation}
\end{figure*}

So far we have discussed various wide binary formation channels and how they may or may not induce metallicity and age dependence in the observed wide-binary fraction. In reality, these mechanisms must all play a role in varying degrees. It is unlikely that our result can be explained by only one mechanism. In particular, no one formation mechanism can explain the non-monotonic relation between metallicity and the wide-binary fraction.

Fig.~\ref{fig:explanation} summarizes the metallicity dependence of the wide binary formation channels, and we propose that the observed metallicity and age dependence are caused by the combination these formation mechanisms. Briefly, the higher stellar density in the star formation environments and the dissolution of higher-mass clusters at an earlier time result in the lower wide-binary fraction in the older stars and the positive metallicity correlation at [Fe/H]$\lesssim0$. The metallicity dependence of dynamical unfolding of compact triples follows the anti-correlation between metallicity and the close-binary fraction, which may explain the declining wide-binary fraction at the super-solar metallicity regime. The values and slopes of the lines in Fig.~\ref{fig:explanation} are only for schematic illustration.

The reason that the positive metallicity correlation of wide-binary fraction ceases at [Fe/H]$=0$ may be that the mean stellar age is similar at [Fe/H]$=0$ and $=0.5$ \citep{Casagrande2011,Bensby2014,SilvaAguirre2018}. Since their formation times are similar, there is no much difference in their formation environments. As the environmental effect ceases to play a dominant role, the anti-correlation with metallicity inherited from the close-binary fraction manifests itself at the regime of super-solar metallicity.

If the negative correlation with metallicity is due to the dynamical unfolding of compact triples, it is not directly obvious why it only applies to [Fe/H]$>0$ and does not extend to [Fe/H]$<0$. One possibility is that at [Fe/H]$<0$, dense environments disrupt the wide binaries or prohibit their formation in the first place. Another challenge for the dynamical unfolding of compact triples to explain the metallicity dependence at [Fe/H]$>0$ is that both the wide-binary fraction and the close-binary fraction \citep{Moe2019} decreases by a factor of 2 from [Fe/H]$=0$ to [Fe/H]$=0.5$. If the metallicity dependence of the wide-binary fraction is inherited from the close-binary fraction, it implies that nearly all wide binaries at [Fe/H]$>0$ are associated with the close binary formation. Therefore, it is possible that there are other mechanisms, like radial migration explained below, that also shape the metallicity dependence at [Fe/H]$>0$.

Given that the wide-binary fraction conspicuously peaks around the solar metallicity, close to the current metallicity of the interstellar medium in the solar neighborhood, it is natural to speculate if the metallicity trend in the wide-binary fraction is due to the radial migration of stars in the Milky Way. Stars that do not have solar metallicities were preferentially formed elsewhere and then radially migrated to the solar neighborhood (e.g. \citealt{Wojno2016,Minchev2018,Han2020}). If the radial migration process can disrupt wide binaries, it would result in a lower wide-binary fraction at low and high [Fe/H]. For example, if a wide binary is trapped at the corotation resonance of a spiral arm, there could be a close destructive interaction between the wide binary and a high-density clump. The migration timescales across the disk are several Gyr (e.g. \citealt{Frankel2018,Frankel2020}), which may explain the inferred age dependence of the wide-binary fraction in Fig.~\ref{fig:metallicity-vtot}. However, as discussed in Sec.~\ref{sec:disruption}, the disruption timescale of a 1000-AU wide binary is longer than the age of the Universe, so we consider the disruption by radial migration processes unlikely to explain the metallicity dependence. 


Even if radial migration processes do not directly disrupt wide binaries, radial migration may still play a role in shaping the metallicity dependence of the wide-binary fraction. For stars with super-solar metallicities in the solar neighborhood, they were formed in the inner Milky Way and then migrated to their current location \citep{Kordopatis2015,Wojno2016, Han2020}. The higher stellar density at the inner Milky Way lowers the wide-binary fraction, and therefore we would expect a lower wide-binary fraction for stars with higher super-solar metallicities. Similarly, stars with sub-solar metallicities may have a wide-binary fraction different from that of solar-metallicity stars due to radial migration. If there is a higher probability of radial migration for stars with more circular orbits (i.e. populations with cooler kinematics), as proposed by \cite{Daniel2018}, then the derived age estimates for this population of radial migrators will be biased low. We include radial migration in Fig.~\ref{fig:explanation}, and future work is needed to determine the relative importance of the scenarios listed in Fig.~\ref{fig:explanation}.

Finally, for the entire population (irrespective of metallicity), the wide-binary fraction of the thick-disk stars is lower than that of the thin-disk stars, and that of the halo stars is marginally lower than the thick-disk stars (Fig.~\ref{fig:metallicity-population}). However, the age distribution of the thin-disk stars is different from the thick-disk and halo stars. \cite{SilvaAguirre2018} show that the age distribution of the low-$\alpha$-element disk (thin disk) peaks at $2$\,Gyr, while the high-$\alpha$-element disk (thick disk) peaks at $11$\,Gyr. Halo stars in the solar neighborhood are also $11$\,Gyr old (e.g. \citealt{Jofre2011,Kalirai2012}). Therefore, the lower wide-binary fraction in the thick-disk and halo stars may be due to that they are older than the thin-disk stars. This suggests that their wide-binary fractions are likely driven by the same effect as the thin-disk stars, which is mostly determined by the formation environments at the different time.

In this study, we propose that multiple formation mechanisms are responsible for the metallicity and age dependence of the wide-binary fraction. Several lines of future work may be able to further constrain their individual contributions. First, different formation mechanisms predict different mass-ratio distributions. For example, the mass ratio distribution from cluster dissolution is consistent with random pairing \citep{Kouwenhoven2010}, while that of the dynamical unfolding of compact triples is not \citep{Reipurth2012}. Therefore, an investigation in the mass ratios of wide binaries as a function of metallicity may shed light on the underlying formation mechanisms. Second, a statistical study of wide binary eccentricity (e.g. \citealt{Tokovinin2016}) as a function of metallicity may be helpful, because dynamical unfolding of compact triples leads to more eccentric outer orbits. However, the interpretation may be complicated, because multiple mechanisms may be at work at the same time, and these wide companions from dynamical unfolding may interact with their formation environments, altering their eccentricity. Also, spectroscopic age estimates for giants via C/N-related features in spectra (e.g. \citealt{Martig2016,Ting2019a}) can further constrain the age evolution of wide binaries.

\section{Conclusions}
\label{sec:conclusion}

In this paper, we investigate the metallicity and age dependence of the wide-binary ($a=10^3$-$10^4$\,AU) fraction. Specifically, we use the metallicity and radial velocity from LAMOST DR5 combined with the astrometric information from Gaia DR2 to measure the wide-binary fraction of field F and G dwarfs. Our findings include: 

\begin{enumerate}
	\item Wide-binary fraction strongly depends on the metallicity (Fig.~\ref{fig:metallicity}). As metallicity increases, wide-binary fraction first increases, peaks at [Fe/H]$\simeq 0$, and then decreases at the high metallicity end. The wide-binary fraction at [Fe/H]$=0$ is about two times larger than that at [Fe/H]$=-1$ and [Fe/H]$= +0.5$. Such metallicity dependence is dominated by the thin-disk stars (Fig.~\ref{fig:metallicity-population}). 
	
	\item The wide-binary fraction is further dependent on the stellar age, with younger stars having a higher wide-binary fraction (Fig.~\ref{fig:metallicity-v}, Fig.~\ref{fig:metallicity-vtot}).
	
	\item Our results suggest that multiple formation channels may be responsible for the formation of wide binaries, resulting in the metallicity and age dependence of the wide-binary fraction (Fig.~\ref{fig:explanation}). Binaries of $10^3$-$10^4$\,AU are unlikely to be disrupted by the gravitational perturbations on the relevant timescale. The positive correlation between the wide-binary fraction and metallicity at [Fe/H]$<0$ may be due to that the density of formation environments and the cluster masses are higher at earlier times, lowering the wide-binary fraction at the low-metallicity end. This also explains the age dependence that younger stars have a higher wide-binary fraction. The anti-correlation between metallicity and the wide-binary fraction at [Fe/H]$>0$ can be inherited from the similar anti-correlation of the close-binary fraction through the dynamical unfolding of compact triples. Radial migration may also enhance the wide-binary fraction around the solar metallicity in the solar neighborhood. 
\end{enumerate}

The authors are grateful to the anonymous referee and Kareem El-Badry for their constructive comments which helped improve the paper. HCH thanks Maosheng Xiang for the discussion on the comparison of LASP and DD-Payne metallicity, Yueh-Ning Lee for the discussion on the metallicity dependence of the formation environments, and David Nataf on the discussion about globular clusters. HCH appreciates the helpful discussion with Jacob Hamer on the manuscript. RFGW thanks her sister, Katherine Barber, for her support and acknowledges the generosity of Eric and Wendy Schmidt, through the recommendation of the Schmidt Futures Program. YST is grateful to be supported by the NASA Hubble Fellowship grant HST-HF2-51425.001 awarded by the Space Telescope Science Institute. HCH is supported in part by the NASA ADAP grant and by Space@Hopkins.

This research makes use of Astropy,\footnote{http://www.astropy.org} a community-developed core Python package for Astronomy \citep{astropy2013, astropy2018}.

\section{Data availability}
The data underlying this article are available in the article and in its online supplementary material.

\appendix

\section{Tests of different selection criteria}
\label{sec:appendix-test}
\begin{figure*}
	\centering
	\includegraphics[height=.45\linewidth]{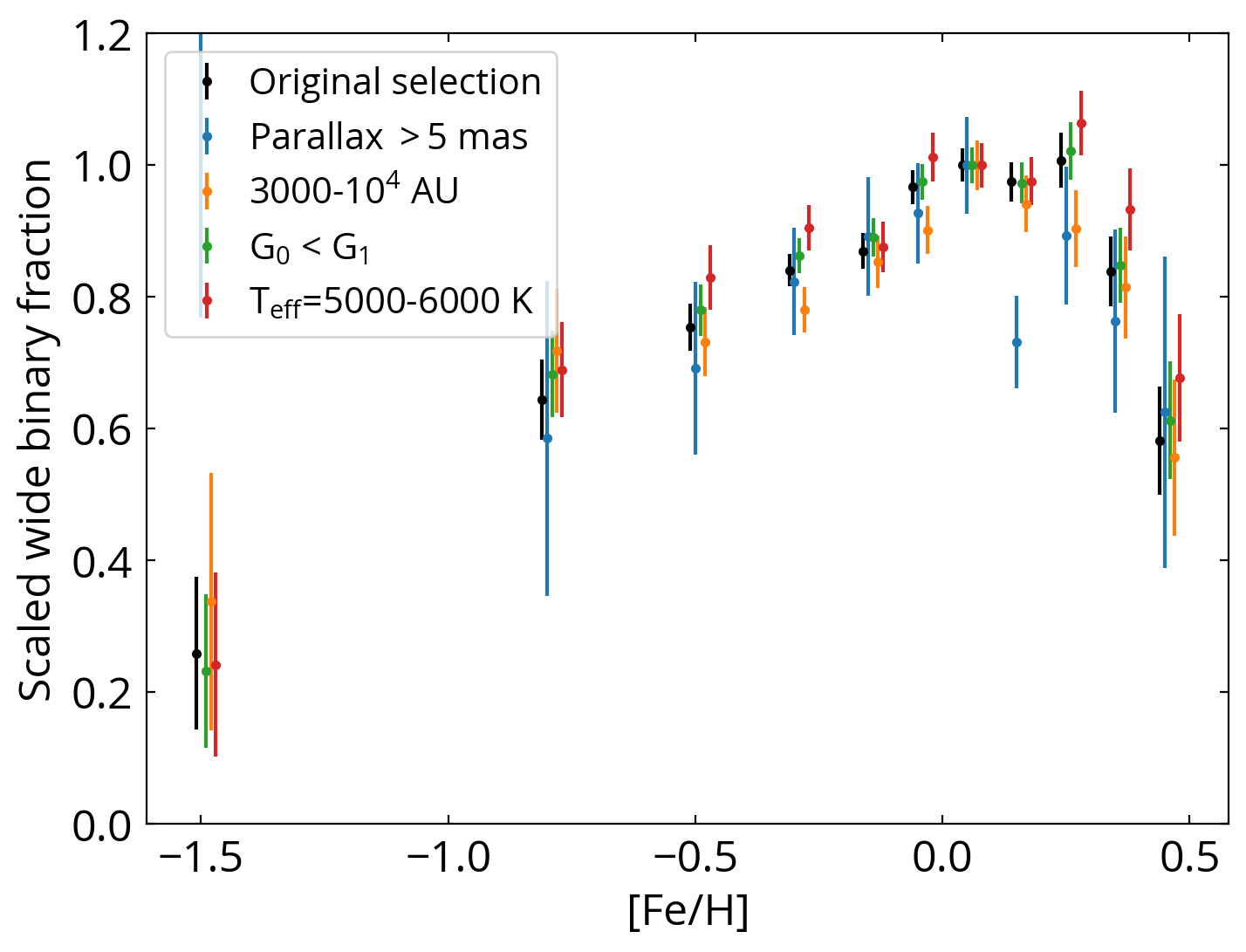}
	
	\caption{The test of different selection criteria. The black points show the original result from Fig.~\ref{fig:metallicity}, and other colors show the results when one certain selection criterion is changed. The points are offset horizontally for clarity. These tests agree well with our main result, supporting that our conclusions are robust against different selection details. }
	\label{fig:test}
\end{figure*}

In Fig.~\ref{fig:test}, we investigate how our results depend on the selection criteria. Based on the selection described in Sec.~\ref{sec:main-selection}, we change a certain criterion and check how it affects the resulting metallicity dependence of the wide-binary fraction. The LASP metallicity is used. The black points are the main LASP result in Fig.~\ref{fig:metallicity}. For better visual comparison, we present the wide-binary fraction scaled to 1 at [Fe/H]$=0$ for the vertical axis in Fig.~\ref{fig:test}. Different tests are slightly offset along the horizontal axis for clarity. The metallicity bins are the same as in Fig.~\ref{fig:metallicity}. We change the following criteria for each test. (1) We select sample with parallax $>5$\,mas (blue), i.e. distances within 200\,pc. (2) Instead of wide binary separations between $10^3$ to $10^4$\,AU, we only consider those with separations between 3000\,AU and $10^4$\,AU (orange). These wide binaries have angular separations $>6$\,arcsec. (3) Since (wide) binary properties depend on the primary mass, we require that the LAMOST stars be the primary of the wide binaries, i.e. the G-band magnitude of the LAMOST star is brighter than that of the companion star ($G_0<G_1$, green). (4) A narrower temperature range of 5000-6000\,K is used (red) to investigate the mass dependence across the entire metallicity range. 



All the tests in Fig.~\ref{fig:test} have a similar metallicity trend as in Fig.~\ref{fig:metallicity}, supporting that our conclusions are robust against the selection details and other potential systematics. Test (1) shows that a similar metallicity trend can already be seen with a sample within 200\,pc, with much larger errors. This result emphasizes the need for a larger sample out to 500\,pc. The binary angular separations in test (2) are $>6$\,arcsec, implying that our conclusion is not affected by the reduced source completeness at small separations. Test (3) shows that the metallicity dependence is nearly the same when we require that the LAMOST stars are the primaries of the wide binaries. The result of test (4) remains similar when a narrower temperature (and therefore mass) range is used, meaning that the metallicity dependence of the wide-binary fraction is not due to the different mass distribution across the metallicity.


\section{Catalogs of wide binaries}
\label{sec:appendix-catalog}
We provide two catalogs electronically for wide binaries with LASP and DD-Payne metallicities, respectively. These wide binaries are the sample used in Fig.~\ref{fig:metallicity}, and they follow the selection criteria detailed in Sec.~\ref{sec:main-selection}. Table~\ref{tab:catalog} tabulates the descriptions for the catalogs. Fields starting with the prefix `0\_' are the information for the LAMOST stars, and those starting with the prefix `1\_' are for the wide companions. Pairs where both stars were observed by LAMOST have two entries in the catalogs.

\begin{table*}[]
	\centering
	\caption{Descriptions for the wide binary catalogs.}
	\label{tab:catalog}
	\begin{tabular}{ll} 
		\hline \hline
		Field 							& Description \\
		\hline
		\texttt{0\_source\_id}    		  & Gaia DR2 source\_id of the LAMOST star \\
		\texttt{0\_ra} 						    & Right ascension of the LAMOST star from Gaia DR2 (J2015.5; deg) \\
		\texttt{0\_dec} 						   & Declination of the LAMOST star from Gaia DR2 (J2015.5; deg) \\
		\texttt{0\_parallax} 				& Parallax of the LAMOST star from Gaia DR2 (mas) \\
		\texttt{0\_parallax\_error}       & Uncertainty in \texttt{0\_parallax} from Gaia DR2 (mas) \\
		\texttt{0\_pmra}		 				& Proper motion in right ascension direction of the LAMOST star from Gaia DR2 (mas yr$^{-1}$) \\
		\texttt{0\_pmra\_error}           & Uncertainty in \texttt{0\_pmra} (mas yr$^{-1}$) from Gaia DR2 \\
		\texttt{0\_pmdec}		 				& Proper motion in declination direction of the LAMOST star from Gaia DR2 (mas yr$^{-1}$) \\
		\texttt{0\_pmdec\_error}           & Uncertainty in \texttt{0\_pmdec} (mas yr$^{-1}$) from Gaia DR2 \\
		\texttt{0\_g}							& Apparent G-band magnitude of the LAMOST star from Gaia DR2 (mag) \\
		\texttt{0\_designation}\tablenotemark{a}			&  LAMOST designation from the LASP catalog \\ 
		\texttt{0\_starid}\tablenotemark{b}					&  Star ID from the DD-Payne catalog \\ 
		\texttt{0\_feh}                    & Iron abundance of the LAMOST star measured by LASP or DD-Payne (dex) \\
		\texttt{0\_teff} 					   & Effective temperature of the LAMOST star measured by LASP or DD-Payne (K) \\
		\texttt{0\_logg} 					  & Surface gravity of the LAMOST star measured by LASP or DD-Payne ($\log$ cgs) \\
		\texttt{0\_vtot}						   & Total velocity ($v_{tot}$) with respect to the local standard of rest (\kms) \\
		\texttt{0\_zmax}						& The maximum Galactic height of the Galactic orbits, $z_{max}$ (kpc) \\
		\texttt{1\_source\_id}    		  & Gaia DR2 source\_id of the companion star \\
		\texttt{1\_ra} 						    & Right ascension of the companion star from Gaia DR2 (J2015.5; deg) \\
		\texttt{1\_dec} 						   & Declination of the companion star from Gaia DR2 (J2015.5; deg) \\
		\texttt{1\_parallax} 				& Parallax of the companion star from Gaia DR2 (mas) \\
		\texttt{1\_parallax\_error}       & Uncertainty in \texttt{1\_parallax} from Gaia DR2 (mas) \\
		\texttt{1\_pmra}		 				& Proper motion in right ascension direction of the companion star from Gaia DR2 (mas yr$^{-1}$) \\
		\texttt{1\_pmra\_error}           & Uncertainty in \texttt{1\_pmra} (mas yr$^{-1}$) from Gaia DR2 \\
		\texttt{1\_pmdec}		 				& Proper motion in declination direction of the companion star from Gaia DR2 (mas yr$^{-1}$) \\
		\texttt{1\_pmdec\_error}           & Uncertainty in \texttt{1\_pmdec} (mas yr$^{-1}$) from Gaia DR2 \\
		\texttt{1\_g}							& Apparent G-band magnitude of the companion star from Gaia DR2 (mag) \\
		\texttt{separation}                  & Physical separation of the wide binary (AU) \\
		\texttt{rel\_vel}						& Relative velocity of the wide binary projected on the sky (\kms) \\
	    \hline \hline
	    \multicolumn{2}{l}{
	    	\begin{minipage}{6.5in}
	    		$^a$ This field is only in the LASP wide binary catalog.
	    		$^b$ This field is only in the DD-Payne wide binary catalog.
	    	\end{minipage}
	    }\\
	\end{tabular}
\end{table*}

\section{Tables of numerical data}
\label{sec:appendix-data}

We tabulate the numerical data used in Fig.~\ref{fig:metallicity}, \ref{fig:metallicity-population}, and \ref{fig:metallicity-v} in Table~\ref{tab:fig3}, \ref{tab:fig4l}, \ref{tab:fig4r}, and \ref{tab:fig5}.

\begin{table*}[]
	\caption{Numerical data for Fig.~\ref{fig:metallicity}}
	\label{tab:fig3}
	\begin{tabular}{ccccccc}
		\hline \hline
		[Fe/H] bin & (-2.0, -1.0) & (-1.0, -0.6) & (-0.6, -0.4) & (-0.4, -0.2) & (-0.2, -0.1) & (-0.1, 0.0) \\
		\hline
		LASP & 0.85$\pm$0.38\%  & 2.11$\pm$0.20\%  & 2.47$\pm$0.12\%  & 2.75$\pm$0.08\%  & 2.84$\pm$0.09\%  & 3.16$\pm$0.08\% \\
		& (5/590)  & (113/5364)  & (434/17599)  & (1174/42705)  & (1015/35699)  & (1418/44847) \\
		\hline
		DD-Payne & 1.28$\pm$0.43\%  & 2.38$\pm$0.16\%  & 2.63$\pm$0.10\%  & 2.81$\pm$0.07\%  & 3.12$\pm$0.08\%  & 3.10$\pm$0.08\% \\
		& (9/704)  & (209/8765)  & (653/24872)  & (1595/56745)  & (1388/44481)  & (1440/46499)  \\
		\hline \hline 
		[Fe/H] bin & (0.0, 0.1) & (0.1, 0.2) & (0.2, 0.3) & (0.3, 0.4) & (0.4, 0.5) & \\
		\hline
		LASP & 3.27$\pm$0.08\% & 3.19$\pm$0.10\% & 3.29$\pm$0.14\% & 2.74$\pm$0.17\% & 1.90$\pm$0.27\% & \\
		& (1535/46935)  & (1079/33854)  & (593/17998)  & (246/8968)  & (50/2626) & \\
		\hline
		DD-Payne & 3.26$\pm$0.10\% & 2.92$\pm$0.13\% & 2.64$\pm$0.17\% & 2.30$\pm$0.25\% & 0.90$\pm$0.45\% & \\
		& (1104/33853)  & (527/18056)  & (255/9659)  & (82/3561)  & (4/442) & \\
		\hline \hline     
	\end{tabular}
\end{table*}

\begin{table*}[]
	\caption{Numerical data for Fig.~\ref{fig:metallicity-population}, left}
	\label{tab:fig4l}
	\begin{tabular}{ccccccc}
		\hline \hline
		[Fe/H] bin & (-2.0, -1.0) & (-1.0, -0.6) & (-0.6, -0.4) & (-0.4, -0.2) & (-0.2, -0.1) & (-0.1, 0.0) \\
		\hline
	    $v_{\rm tot}<120$\kms & 0.98$\pm$0.69\% & 2.15$\pm$0.23\% & 2.50$\pm$0.12\% & 2.76$\pm$0.08\% & 2.85$\pm$0.09\% & 3.17$\pm$0.08\%  \\
		& (2/204)  & (87/4038)  & (414/16562)  & (1161/42136)  & (1011/35534)  & (1418/44723)   \\
		\hline \hline 
		[Fe/H] bin & (0.0, 0.1) & (0.1, 0.2) & (0.2, 0.3) & (0.3, 0.4) & (0.4, 0.5) & \\
		\hline
		$v_{\rm tot}<120$\kms & 3.27$\pm$0.08\% & 3.19$\pm$0.10\% & 3.30$\pm$0.14\% & 2.75$\pm$0.18\% & 1.91$\pm$0.27\% & \\
		& (1533/46838)  & (1078/33785)  & (593/17966)  & (246/8950)  & (50/2622) & \\
		\hline \hline 
		[Fe/H] bin & (-2.0, -1.0) & (-1.0, -0.6) & (-0.6, -0.3) & (-0.3, 0.0) & (0.0, 0.5) \\
		\hline
		120-250\kms & 0.70$\pm$0.49\% & 1.94$\pm$0.39\% & 1.99$\pm$0.38\% & 1.91$\pm$0.60\% & 1.38$\pm$0.79\% \\
		& (2/287)  & (25/1290)  & (27/1359)  & (10/524)  & (3/218) \\
		\hline
		$>250$\kms & 1.01$\pm$1.01\% & 2.78$\pm$2.78\%\\
		& (1/99)  & (1/36) \\
		\hline \hline     
	\end{tabular}
\end{table*}

\begin{table*}[]
	\caption{Numerical data for Fig.~\ref{fig:metallicity-population}, right}
	\label{tab:fig4r}
	\begin{tabular}{ccccccc}
		\hline \hline
		[Fe/H] bin & (-2.0, -1.0) & (-1.0, -0.6) & (-0.6, -0.4) & (-0.4, -0.2) & (-0.2, -0.1) & (-0.1, 0.0) \\
		\hline
		$z_{\rm max}<1$\,kpc & 0.80$\pm$0.56\% & 2.13$\pm$0.25\% & 2.44$\pm$0.13\% & 2.72$\pm$0.08\% & 2.86$\pm$0.09\% & 3.20$\pm$0.09\%  \\
		& (2/251)  & (72/3383)  & (347/14206)  & (1065/39130)  & (982/34315)  & (1399/43765)   \\
		\hline \hline 
		[Fe/H] bin & (0.0, 0.1) & (0.1, 0.2) & (0.2, 0.3) & (0.3, 0.4) & (0.4, 0.5) & \\
		\hline
		$z_{\rm max}<1$\,kpc & 3.28$\pm$0.08\% & 3.20$\pm$0.10\% & 3.32$\pm$0.14\% & 2.78$\pm$0.18\% & 1.94$\pm$0.27\% \\
		& (1510/46041)  & (1067/33339)  & (587/17707)  & (245/8824)  & (50/2578) & \\
		\hline \hline 
		[Fe/H] bin & (-2.0, -1.0) & (-1.0, -0.6) & (-0.6, -0.3) & (-0.3, 0.0) & (0.0, 0.5) \\
		\hline
		1-5\,kpc &1.11$\pm$0.64\% & 2.07$\pm$0.33\% & 2.76$\pm$0.23\% & 2.45$\pm$0.24\% & 2.33$\pm$0.35\% \\
		& (3/270)  & (40/1931)  & (144/5211)  & (103/4196)  & (44/1889) \\
		\hline
		$z_{\rm max}>5$\,kpc & $<1.44$\% & 2.00$\pm$2.00\%\\
		& (0/69)  & (1/50) \\
		\hline \hline     
	\end{tabular}
\end{table*}

\begin{table*}[]
	\caption{Numerical data for Fig.~\ref{fig:metallicity-v}}
	\label{tab:fig5}
	\begin{tabular}{ccccccc}
		\hline \hline
		[Fe/H] bin & (-2.0, -1.0) & (-1.0, -0.6) & (-0.6, -0.4) & (-0.4, -0.2) & (-0.2, -0.1) & (-0.1, 0.0) \\
		\hline
		$v_{\rm tot}<30$\kms & $<11$\% & 1.59$\pm$0.79\% & 2.28$\pm$0.32\% & 2.83$\pm$0.17\% & 3.22$\pm$0.17\% & 3.33$\pm$0.14\%  \\
		& (0/9)  & (4/252)  & (51/2238)  & (263/9308)  & (375/11633)  & (595/17888)  \\
		\hline
		30-60\kms & 3.23$\pm$3.23\% & 2.52$\pm$0.50\% & 2.53$\pm$0.19\% & 2.85$\pm$0.12\% & 2.73$\pm$0.13\% & 3.25$\pm$0.13\% \\
		& (1/31)  & (25/994)  & (172/6786)  & (550/19287)  & (435/15944)  & (613/18863)  \\
		\hline
		60-120\kms & $<1.15$\% & 1.89$\pm$0.34\% & 2.48$\pm$0.23\% & 2.41$\pm$0.15\% & 2.56$\pm$0.20\% & 2.76$\pm$0.20\% \\
		& (0/87)  & (31/1636)  & (117/4724)  & (246/10218)  & (170/6635)  & (191/6928) \\
		\hline \hline 
		[Fe/H] bin & (0.0, 0.1) & (0.1, 0.2) & (0.2, 0.3) & (0.3, 0.4) & (0.4, 0.5) & \\
		\hline
		$v_{\rm tot}<30$\kms & 3.59$\pm$0.13\% & 3.62$\pm$0.16\% & 4.06$\pm$0.24\% & 2.81$\pm$0.30\% & 2.64$\pm$0.59\% \\
		&  (736/20481)  & (525/14509)  & (291/7174)  & (85/3027)  & (20/757) \\
		\hline
		30-60\kms & 3.15$\pm$0.13\% & 2.89$\pm$0.14\% & 2.86$\pm$0.19\% & 3.00$\pm$0.27\% & 1.52$\pm$0.35\% \\
		& (599/18991)  & (405/14022)  & (220/7680)  & (121/4040)  & (19/1246) \\
		\hline
		60-120\kms & 2.66$\pm$0.20\% & 2.86$\pm$0.25\% & 2.68$\pm$0.31\% & 2.23$\pm$0.36\% & 1.92$\pm$0.58\% \\
		& (173/6499)  & (136/4753)  & (76/2831)  & (39/1745)  & (11/572) \\
		\hline \hline     
	\end{tabular}
\end{table*}

\bibliography{paper-WideBinary}{}
\bibliographystyle{aasjournal}
\end{document}